\documentclass[12pt]{article}
\usepackage{amsmath,amsfonts,amsbsy}
\usepackage[dvips]{graphicx,epsfig}
\usepackage[vcentermath,enableskew]{youngtab}

\def\hybrid{\topmargin -20pt    \oddsidemargin 0pt
        \headheight 0pt \headsep 0pt
        \textwidth 6.25in       
        \textheight 9.5in       
        \marginparwidth .875in
        \parskip 5pt plus 1pt   \jot = 1.5ex}

\hybrid

\renewcommand{\theequation}{\thesection.\arabic{equation}} \csname
@addtoreset\endcsname{equation}{section}

\newcommand{\hoch}[1]{$^{#1}$}
\newcommand{\bein}[3]{{#1}_{#2}^{\hspace{0.5em}{#3}}}

\newcommand{\spinleg}[4]{{#1}_{#2\hspace{0.7em}#4}^{\hspace{0.5em}{#3}}}

\def\moth{\mathsurround=0pt}
\newdimen\zo \zo=0pt

\def\tick{\leaders\hrule height 0.5ex depth 0pt \hskip 0.5pt}
\def\upboxfill{$\moth \setbox\zo\hbox{\tick}%
  \hskip 3pt\hbox to 0pt{$\tick$\hss}\hrulefill \hbox to 7.5pt{$\tick$\hss}$}
\def\underbox#1{\offinterlineskip{\mathord{\mathop{\vtop{\moth\ialign{##\crcr
      $\hfil\displaystyle{#1}\hfil$\crcr\noalign{}
      {\upboxfill}\crcr\noalign{}}}}\limits}}}
\def\dtick{\leaders\hrule height .34pt depth 0.5ex \hskip 0.5pt}
\def\downboxfill{$\moth \setbox\zo\hbox{\dtick}%
  \hskip 2pt\hbox to 0pt{$\dtick$\hss}\hrulefill \hbox to 2pt{$\dtick$\hss}$}

\def\undersym#1{\underbox{{}#1}}

\def\ystar{{*}}

\def\mso{\mathfrak{so}}

\def\msp{\mathfrak{sp}}
\def\mosp{\mathfrak{osp}}

\def\mhso{\mathfrak{ho}}

\def\bec{\begin{center}}
\def\ec{\end{center}}
\def\a{\alpha}  

\def\b{\beta}

\def\d{\delta} 
\def\D{\Delta}
\def\e{\epsilon}

\def\l{\lambda}
\def\L{\Lambda}

\def\n{\nu}
\def\r{\rho}
\def\s{\sigma}

\def\cB{{\cal B}}

\def\cO{{\cal O}}
\def\cA{{\cal A}}

\def\cU{{\cal U}}

\def\cI{{\cal I}}

\def\cA{{\cal A}}

\def\cO{{\cal O}}

\def\del{\partial}

\let\la=\label

\def\nn{\nonumber}
\newcommand{\eq}[1]{(\ref{#1})}

\def\tr{{\rm tr}}
\def\Tr{{\rm Tr}}

\def\be{\begin{equation}}
\def\ee{\end{equation}}
\def\bea{\begin{eqnarray}}
\def\eea{\end{eqnarray}}
\def\ba{\begin{array}}
\def\ea{\end{array}}

\def\ft#1#2{{\textstyle{{\scriptstyle #1}
\over {\scriptstyle #2}}}}

\def\scs#1{\section{\sc #1}}
\def\scss#1{\subsection{\sc #1}}

\begin{document}

\hfill{\texttt{ITP-UU-07/29}}

\vspace{-5pt}

\hfill{\texttt{SPIN-07/19}}

\vspace{-5pt}


\hfill{May 2007}

\vspace{20pt}

\begin{center}


{\Large\sc Higher-Spin Chern-Simons Theories \\[0.15cm] in odd
Dimensions}


\vspace{30pt}
{\sc Johan Engquist\hoch1 and  Olaf Hohm\hoch1}\\[15pt]

\hoch{1}{\it\small Institute for Theoretical Physics and Spinoza
Institute, Utrecht University \\3508 TD Utrecht, The Netherlands}


\vspace{35pt}  \end{center}

\begin{center} {\bf ABSTRACT}\\[3ex]

\begin{minipage}{13cm}
\small We construct consistent bosonic higher-spin gauge theories
in odd dimensions $D>3$ based on Chern-Simons forms. The gauge
groups are infinite-dimensional higher-spin extensions of the
Anti-de Sitter groups $SO(D-1,2)$. We propose an invariant tensor
on these algebras, which is required for the definition of the
Chern-Simons action. The latter contains the purely gravitational
Chern-Simons theories constructed by Chamseddine, and so the
entire theory describes a consistent coupling of higher-spin
fields to a particular form of Lovelock gravity. It contains
topological as well as non-topological phases. Focusing on $D=5$
we consider as an example for the latter an $AdS_4\times S^1$
Kaluza-Klein background. By solving the higher-spin torsion
constraints in the case of a spin-3 field, we verify explicitly
that the equations of motion reduce in the linearization to the
compensator form of the Fr\o nsdal equations on $AdS_4$.
\end{minipage}
\end{center}

 {\vfill\leftline{}\vfill \vskip  10pt \footnoterule
{\footnotesize E-mails: \texttt{J.Engquist@phys.uu.nl, O.Hohm@phys.uu.nl} \vskip
-12pt}}

\setcounter{page}{1}

\pagebreak

\tableofcontents

\scs{Introduction} The construction of theories describing
consistently interacting higher-spin fields is for several reasons
of great interest. For one thing string theory contains an
infinite tower of massive higher-spin states, and it is an old
idea that these hint to a spontaneously broken phase of a theory
with a huge hidden gauge symmetry, thus extending the geometrical
framework of Einstein's theory
\cite{Gross:1988ue,Isberg:1993av,Sundborg:2000wp,Bonelli:2003kh,Sagnotti:2003qa,Engquist:2005yt}.
However, the actual formulation of higher-spin theories is usually
precluded by the interaction problem. The latter refers to the
apparent impossibility of introducing interactions into a free
higher-spin (HS) theory in such a way, that the number of
dynamical degrees of freedom is unaltered
\cite{Sorokin:2004ie,Weinberg:1980kq}. For instance, naively
coupling free massless HS fields to gravity violates the HS gauge
symmetry and thus renders the theory inconsistent
\cite{Aragone:1979bm}.

In a series of paper Vasiliev has, however, begun to find a route
avoiding these no-go theorems, i.e.~to consistently couple HS
fields to gravity, by relaxing the following assumptions. First,
the theory is assumed to have a non-vanishing negative
cosmological constant -- leading to Anti-de Sitter ($AdS$) instead
of Minkowski space as the `ground state' -- and to depend on this
cosmological constant in a non-polynomial way. The latter excludes
a flat space limit, in accordance with standard S-matrix arguments
\cite{Coleman:1967ad}. Second, it will necessarily contain an
infinite number of massless fields carrying arbitrarily high spin,
whose couplings can be of arbitrary power in the derivatives (see
\cite{Bekaert:2005vh} for a review and references therein).

The formulation of the associated HS theory is based on a gauging
of an infinite-dimensional HS algebra, in the same way that
gravity and supergravity theories can be viewed as resulting from
a gauging of a (super-)$AdS$ algebra. However, theories which are
constructed along these lines (as, e.g., in the approach of
MacDowell-Mansouri \cite{MacDowell:1977jt}) are not true gauge
theories in that the gauge symmetry is not manifest and, moreover,
(super-)torsion constraints have to be imposed by hand. For
instance, in supergravity invariance under local supersymmetry is
by no means manifest and has to be checked explicitly. In
addition, so-called `extra fields' appear in HS theories, which
are unphysical and have to be expressed in terms of the physical
fields by imposing further constraints. In total, the program of
Vasiliev consists of finding a non-linear HS theory
\cite{Bekaert:2005vh}, which
 \begin{itemize}
  \item[(i)]
   is still invariant under (a deformation of) the HS
   symmetry, and
  \item[(ii)]
   yields in the linearization the required free field
   equations.
 \end{itemize}
Of course, both requirements are related since once it is proven
that the free field equations are of 2$^{\rm nd}$ order, the HS
symmetry, i.e.~(i), fixes the field equations uniquely to the
so-called Fr\o nsdal form. In the approach of
\cite{Fradkin:1986qy,Vasiliev:2001wa} this requirement is
implemented through the condition that the extra fields decouple
in the free limit (for reasons we will explain below). However,
these conditions have no natural interpretation from the point of
view of the HS gauge symmetry. In turn the consistency of the
resulting HS action can only be checked up to some order, as it
has been done in $D=4$ and $D=5$ for cubic couplings
\cite{Fradkin:1986qy,Vasiliev:2001wa}. But there are even reasons
to expect that this consistency will not extend to all orders
\cite{Vasiliev:2001wa}. In fact, up to date a fully consistent
action describing interactions of propagating HS fields is not
known.

An approach, which is instead followed in order to describe
consistent HS interactions at the level of the equations of
motion, is given by the so-called `unfolded formulation'
\cite{Vasiliev:2005zu,Vasiliev:1992gr,Vasiliev:1989xz}. The latter
is a surprisingly concise way to keep the HS invariance manifest.
However, in this approach there is not only an infinite number of
physical HS fields, but each of the infinite fields has an
infinite number of auxiliary fields, which, roughly speaking,
parametrize all space-time derivatives of the physical field. This
in turn complicates the analysis of the physical content, and it
would be clearly desirable to have a conventional action principle
that extends the Einstein-Hilbert action in the same way as
supergravity does for spin-3/2 fields.

Concerning the problem of finding a consistent HS action, it
should be noted that one example does exist: the Chern-Simons
action in $D=3$ constructed by Blencowe based on a HS algebra
\cite{Blencowe:1988gj}. (See also
\cite{Fradkin:1989xt,Bergshoeff:1989ns} and
\cite{Hohm:2005sc,Hohm:2006ud} in a related context.) As the
Chern-Simons theory is a true gauge theory, the resulting HS
theory is consistent by construction and naturally extends the
Einstein-Hilbert action (which in $D=3$ also has an interpretation
as a Chern-Simons action \cite{Witten:1988hc}.) It is, however,
only of limited use since it is topological and does not give rise
to propagating degrees of freedom. On the other hand, gauge
invariant Chern-Simons actions exist in all odd dimensions, and
even though they are topological in any dimension in the sense
that they do not depend on a metric, they are not devoid of local
dynamics in $D>3$. In fact, it has been shown by Chamseddine
\cite{Chamseddine:1989nu,Chamseddine:1990gk} that the Chern-Simons
actions based on the $AdS_D$ algebras $\mso(D-1,2)$ are equivalent
to a particular type of Lovelock gravity with propagating torsion
and thus by far not dynamically trivial. So one might wonder what
happens if one defines a Chern-Simons action based on a HS
extension of $\mso(D-1,2)$. This paper is devoted to the analysis
of this question.

The organization of the paper is as follows. In sec.~2 we briefly
review the known free HS theories on Minkowski and $AdS$, and we
introduce the HS Lie algebras which will later on serve as gauge
algebras. The general construction of Chern-Simons actions in odd
dimensions will be reviewed in sec.~3.1, together with the
realization of Lovelock gravity as a Chern-Simons gauge theory. In
sec.~3.2 we construct an invariant tensor of the HS algebra, which
in turn allows a consistent extension of Chern-Simons gravity to
include an infinite tower of HS fields. The constructed theory is
then linearized around the `non-topological' Kaluza-Klein
background $AdS_4\times S^1$ in sec.~4. Focusing on the spin-3
mode, we show that the equations of motion reduce to the correct
free equations on $AdS_4$. We conclude in sec.~5, while technical
details concerning Young tableaux, the symmetric invariant and the
spin-3 Riemann tensor are relegated to the appendices A--C.

\scs{Higher-Spin Theories and Their Gauge Algebras} In this
section we first review free HS theories on Minkowski and $AdS$
backgrounds, and then introduce the infinite-dimensional HS Lie
algebras, which are the starting point for the construction of
interacting HS theories. The results hold in general odd
dimensions, though for concreteness we will often specify to
$D=5$.

\scss{Free Higher-Spin Actions}\label{free} Bosonic fields of
arbitrary spin $s$ are described by symmetric rank-$s$ tensors
$h_{\mu_1...\mu_s}$. In the massless case they are subject to the
gauge symmetry
 \bea
  \delta_{\epsilon}h_{\mu_1...\mu_s}=\nabla_{(
  \mu_1}\epsilon_{\mu_2...\mu_s )}\;,
 \eea
parametrized by a symmetric transformation parameter $\epsilon$ of
rank $s-1$. An action for a free field of spin $s$ on Minkowski
and (anti-)de Sitter backgrounds has been given by Fr\o nsdal
\cite{Fronsdal:1978rb,Fronsdal:1978vb}. For a spin-$3$ field
$h_{\mu\nu\rho}$, which is the case we will later on examine in
more detail, it is of the form
 \bea\label{spinaction}
 \begin{split}
  S\ =\ \frac{1}{2}\int d^D x
  \big[&\nabla_{\mu}h_{\nu\rho\sigma}\nabla^{\mu}h^{\nu\rho\sigma}
  -3\nabla_{\mu}h^{\rho}{}_{\rho\nu}\nabla^{\mu}h_{\sigma}{}^{\sigma\nu}
  +3\cdot2
  \nabla_{\mu}h^{\rho}{}_{\rho\nu}\nabla_{\sigma}h^{\sigma\mu\nu}
  \\
  &-3\nabla^{\mu}h_{\mu\nu\rho}\nabla_{\sigma}h^{\sigma\nu\rho}
  -\frac{3}{2}\nabla_{\mu}h^{\mu\rho}{}_{\rho}\nabla_{\nu}h^{\nu\sigma}{}_{\sigma}
  +{\cal L}_m\big]\;.
 \end{split}
 \eea
Here $\nabla_{\mu}$ denotes the $AdS$-covariant derivative or a
partial derivative in case of a Minkowski background. In the flat
case the additional term ${\cal L}_m$ vanishes, while on $AdS$ the
HS gauge symmetry requires a mass-like term proportional to the
cosmological constant. The latter then amounts to the equations of
motion
 \bea
  {\cal F}_{\mu\nu\rho}^{\rm AdS} \equiv \square h_{\mu\nu\rho} -
  3\nabla_{(\mu}\nabla\cdot h_{\nu\rho)} +
  3\nabla_{(\mu}\nabla_{\nu}h_{\rho )}^{\prime}
  -\frac{1}{L^2}[(D-3)h_{\mu\nu\rho}+2\cdot
  3g_{(\mu\nu}h_{\rho)}^{\prime}]\ =\ 0\;,
 \eea
which defines the so-called Fr\o nsdal operator ${\cal F}$. Here
$h_{\mu}^{\prime}$ denotes the trace of $h$ in the $AdS$ metric
and $L$ is the $AdS$ radius, related to the cosmological constant
by $L=1/\sqrt{\Lambda}$. Let us finally note that the given action
or field equations are invariant under the gauge variations only
if the transformation parameter is traceless and that for spin
$s>3$ a double-tracelessness condition has to be imposed on the
fields in order to give rise to the correct number of spin-$s$
degrees of freedom \cite{deWit:1979pe}.

The difficulty in promoting these HS theories to interacting
theories via coupling to gravity or electrodynamics is due to the
fact that the presence of generic covariant derivatives in
(\ref{spinaction}) violates the HS gauge symmetry. This in turn
implies that the unphysical degrees of freedom are no longer
eliminated and the theory becomes inconsistent. Despite of these
negative results, Vasiliev has pioneered an approach towards a
consistent coupling of HS fields to gravity, which is based on the
introduction of an infinite-dimensional HS algebra
\cite{Fradkin:1986ka}. The latter requires a frame-like
formulation of HS fields, which mimics the vielbein formulation of
general relativity rather than the metric-like formulation used in
(\ref{spinaction}) \cite{Vasiliev:1980as}. More specifically, a
spin-3 field, for instance, is described by $\bein{e}{\mu}{ab}$,
being symmetric in the frame indices $a,b$, together with an
analogue of the spin-connection $\bein{\omega}{\mu}{ab,c}$. A
closer inspection has, however, revealed that consistency of the
HS algebra requires further fields, which are the so-called extra
fields. These issues will be dealt with in later sections, but for
the moment we just note that the resulting algebra will be derived
from the enveloping algebra of the $AdS_D$ algebra $\mso(D-1,2)$,
to which we turn in the next section.

\scss{A Higher-Spin Extension of $\mso(D-1,2)$} \la{sec:ahse} The
starting point for the construction of an infinite-dimensional HS
algebra is the $AdS$ symmetry group $SO(D-1,2)$. The Lie algebra
of the latter is spanned by the anti-hermitian generators
$M_{AB}=-M_{BA}$, $A=0,1,\ldots,D-1,D+1$, obeying the commutation
relations
\begin{eqnarray} \begin{split} \label{so42alg}
[M_{AB},M_{CD}]&\ =\
\eta_{BC}M_{AD}-\eta_{AC}M_{BD}-\eta_{BD}M_{AC}+\eta_{AD}M_{BC}
\;\\ &\ \equiv \ f_{AB,CD}{}^{EF}M_{EF}\ , \end{split}
\end{eqnarray} where \bea \la{structco} \eta_{AB}&=&{\rm
diag}(-1,1,1,1,1,-1 )\ , \qquad f_{AB,CD}{}^{EF}\;=\;4\d_{[A}^{[E}
\eta_{B][C} \d^{F]}_{D]} \ . \eea In the Lorentz basis, the
$\mso(D-1,2)$ commutation relations read \bea \begin{split}
[M_{ab},M_{cd}]&\ =\
\eta_{bc}M_{ad}-\eta_{ac}M_{bd}-\eta_{bd}M_{ac}+\eta_{ad}M_{bc}\ ,
\cr [M_{ab},P_c]&\ =\ -2\eta_{c[a}P_{b]}\ , \qquad
[P_a,P_b]\;=\;M_{ab}\ . \end{split} \eea

To define a Lie algebraic HS extension of $\mso(D-1,2)$ it is
convenient \cite{Vasiliev:2003ev,Sagnotti:2005ns} to introduce a
set of bosonic vector oscillators $y^i_A$, where $i=1,2$ is an
$\msp(2)$ doublet index, obeying an associative non-commutative
star product
 \bea\label{star} && y^i_A\ystar y^j_B\ = \ y^i_Ay^j_B +
\e^{ij}\eta_{AB} \ , \qquad [y^i_A,y^j_B]_{\ystar}\ = \
2\e^{ij}\eta_{AB}\ , \eea where $\e^{ij}=-\e^{ji}$ denotes the
invariant $\msp(2)$ tensor, and we have introduced the bracket
$[U,V]_\ystar= U\ystar V-V\ystar U$. The star product of general
functions $f(y)$ and $g(y)$ can be defined by the Moyal-Weyl
formula \bea \la{moyalvector}
f(y)~\ystar~g(y)&=&\exp\Big(\frac{\del}{\del
y_A^i}\frac{\del}{\del
z_B^j}\e^{ij}\eta_{AB}\Big)f(y)g(z)\Big|_{z=y}\ , \eea which
reduces for linear functions to (\ref{star}).

Given the oscillators we can construct the generators of the
commuting (Howe dual) algebras $\mso(D-1,2)$ and $\msp(2)$ as the
bilinears \bea M_{AB}\ = \ \ft12 y^i_Ay_{iB}\ , \qquad K_{ij} \ =
\ \ft12 y^A_iy_{jA}\ , \eea from which it indeed follows that
$[K_{ij},M_{AB}]_\ystar=0$. The construction of the HS Lie algebra
$\mhso(D-1,2)$ is based on the enveloping algebra  of
$\mso(D-1,2)$. It is defined in terms of the oscillator as
\cite{Vasiliev:2003ev,Sagnotti:2005ns} \bea \la{defhsal}
\mhso(D-1,2)\ = \ \Big\{T(y): ~~T^\dagger=-T,
~~[K_{ij},T]_\ystar=0\Big\}\ , \eea where $T(y)$ are arbitrary
polynomials in the oscillator $y_A^i$, and the last condition
singles out the $\msp(2)$ singlets. This algebra is sometimes
(perhaps misleadingly) referred to as the `off-shell HS algebra'
since it is generated by {\it trace-full generators}. On the other
hand, starting from this algebra one may construct the
corresponding `on-shell algebra' where the generators are made
traceless by factoring out the ideal in $\mhso(D-1,2)$ spanned by
elements of the form $K_{ij}\ystar X^{ij}$
\cite{Vasiliev:2003ev,Sagnotti:2005ns,Bekaert:2005vh}. In the
formulation of HS theories utilizing the approach of unfolded
dynamics \cite{Vasiliev:2005zu,Sezgin:2001zs}, where an action
principle is not needed, the on-shell algebra has been mostly
used. It is only recently \cite{Sagnotti:2005ns} that the
importance of the off-shell algebra has been emphasized. In
contrast, in an ordinary action formulation of HS theories, as the
one in this paper, we believe that an algebra with trace-full
generators is crucial. In the remaining of the paper we will avoid
the term `off-shell algebra'.

The polynomial $T(y)$ appearing in \eq{defhsal} admits a level
decomposition into monomials $T_\ell(y)$ (we associate the
generators $T_\ell$ at level $\ell$ with spins
$s=\ell+2=2,3,4,\ldots$) \bea T(y)\ = \ \sum_{\ell=0}^\infty
T_\ell(y)\ , \qquad T_\ell(\l y)\ = \ \l^{2\ell+2} T_\ell(y)\
.\eea The definition in terms of vector oscillators implies in
particular that the algebra does not contain the full enveloping
algebra spanned by polynomials in $M_{AB}$. Elements which vanish
identically in the vector oscillator formulation belong to a
certain ideal $\cI\subset\cU[\mso(D-1,2)]$. For instance, the
antisymmetric part $M_{[AB}M_{C]D}$ vanishes due to the $\msp(2)$
identity $\epsilon^{[ij}\epsilon^{k]l}=0$. This in turn implies
that the generators of the HS algebra (\ref{defhsal}) are in
specific Young tableaux. In other words, the $T_\ell(y)$ have an
expansion in terms of $GL(D+1)$ tensors \cite{Sagnotti:2005ns}
\bea \la{defgen} T_{A(s-1),B(s-1)}&\equiv& \mathbb P_{(s-1,s-1)}
\Big(M_{A_1B_1}\cdots M_{A_{s-1}B_{s-1}}\Big)\\ & = &
\frac{1}{2^{2s-2}}\mathbb P_{(s-1,s-1)}
\Big(y^{i_1}_{A_1}y_{i_1B_1}\cdots
y^{i_{s-1}}_{A_{s-1}}y_{i_{s-1}B_{s-1}}\Big)\ , \eea where the
$\mso(D-1,2)$ generators appear at level $0$.\footnote{There
exists a further restriction of $\mhso(D-1,2)$ to a minimal
algebra containing only {\it even} spins $s=2,4,6,\ldots$
\cite{Vasiliev:2003ev,Sagnotti:2005ns}.} We have introduced a
notation in which $T_{A(n),B(n)}\equiv T_{A_1\ldots A_n,B_1\ldots
B_n}$, each set of indices being totally symmetrized. $\mathbb
P_{(s-1,s-1)}$ is a Young projector which imposes the symmetry of
the two-row $GL(D+1)$ Young tableau (see appendix \ref{A} for
details) \bea\label{youngbox}
\underbrace{\yng(5,5)\cdots\yng(5,5)}_{s-1}\ ~~ . \eea

Later on we will need the generator in a $GL(D)$ `Lorentz' basis.
Splitting (\ref{youngbox}) accordingly for spin $s$, we find $s$
generators, which schematically are in the tableaux
 \bea
  \yng(4)\;, \qquad \yng(4,1)\;, \qquad \ldots\ ,\qquad \;
  \yng(4,4)\;.
 \eea
More specifically, the generator $T_{A_1...A_{s-1},B_1...B_{s-1}}$
split into the series of generator $T_{a_1...a_{s-1}}$ and
$T_{a_1...a_{s-1},b_1\ldots b_{t}}$ for $1\leq t\leq s-1$. The
gauge fields $\bein{e}{\mu}{a_1...a_{s-1}}$ corresponding to the
first generator will later be identified with the physical
spin-$s$ field, while the fields for the remaining $s-1$
generators are in the literature referred to as the auxiliary
($t=1$) and extra fields ($t>1$). However, for us this distinction
between auxiliary and extra fields will be redundant and we will
therefore henceforth refer to all fields with $t>0$ as auxiliary.

The complete set of commutation relations of the $\mhso(D-1,2)$
algebra is not known in a closed form. Luckily, for the linearized
spin-$s$ analysis, to be treated in sec.~\ref{sec:DA} for an
expansion around a spin-$2$ solution, it is sufficient to specify
the spin-$2$ -- spin-$s$ commutation relations, which are entirely
fixed by the representations theory of the $AdS$ subalgebra
$\mso(D-1,2)$ \bea \la{spintwos}
[M_{AB},T_{C(s-1),D(s-1)}]_*&=&-4(s-1)\mathbb
P_{(s-1,s-1)}\big(\eta\undersym{{}_{AC_{s-1}}T_{\phantom{|}B\phantom{|}}}\hspace{-.1cm}{}_{C(s-2),D(s-1)}\big)\
. \eea Let us finally mention that when commuting a spin-$s$
generator with a spin-$s'$ generator we obtain a sequence of
generators with spins \bea s+s'-2, ~s+s'-4,\ldots,~ |s-s'|+2\ .
\eea Notice that only the $s=2$ subsector is closed.

\scs{Chern-Simons Theories in Odd Dimensions}

In this section we will introduce the formulation of Chern-Simons
theories \cite{Chamseddine:1989nu,Chamseddine:1990gk} in general
odd dimensions (for a review see \cite{Zanelli:2002qm}). The
theory is specified once we give the algebra and the relevant
invariant tensor. We will specify to $AdS$ Lovelock gravity, with
a focus on $D=5$, though the results directly extend to all odd
dimensions (for the explicit formulas see
\cite{Chamseddine:1990gk}.) Although there is no non-trivial
propagation around the vacuum solution $AdS_5$, interestingly, the
theory also admits a simple $AdS_4$ solution
\cite{Chamseddine:1990gk} around which the graviton propagates. In
sec.~\ref{sec:DA} we will analyze the linearized HS dynamics
around this solution. Finally, in this section we will propose an
invariant tensor for the full HS algebra.

\scss{Lovelock Gravity as $\mso(D-1,2)$ Gauge Theory}
\la{sec:love} In any odd dimension $D=2n-1$ a gauge-invariant
Chern-Simons action can be defined, which is based on the
invariant $2n$-form $\langle F^{n}\rangle$ constructed out of the
field strength $F$, with $\langle\hspace{0.4em}\rangle$ denoting
an invariant symmetric tensor of degree $n$. More specifically,
this expression is a total derivative and thus gives rise to a
dynamically non-trivial theory only on the boundary, i.e.~in one
dimension less. The action can be written in closed form as
 \bea\label{CSgauge}
  S_{\rm CS}\ =\ \int_{M_{2n}} \langle F^{n}\rangle\ = \
  \int_{M_{2n-1}} n \int_0^1 dt\langle
  A(tdA+t^2A^2)^{n-1}\rangle\;,
 \eea
where $M_{2n-1}=\partial M_{2n}$, and we left the wedge products
implicit. The resulting Chern-Simons form in $D=2n-1$ is by
construction gauge-invariant up to total derivatives. Explicitly,
one has under arbitrary variations
 \bea\label{CSvar}
  \delta S_{\rm CS}\ =\ n\int_{M_{2n-1}}\langle\delta A\wedge
  F^{n-1}\rangle\;,
 \eea
i.e.~gauge invariance under
$\delta_{\epsilon}A_{\mu}=D_{\mu}\epsilon$ follows by the Bianchi
identity.

For definiteness we focus on $D=5$. The gauge field $A$ then takes
values in the Lie algebra of the group $SO(4,2)$. Specifically, we
write in an $SO(4,1)$ covariant manner
$A_{\mu}=\bein{e}{\mu}{a}P_{a}+\frac{1}{2}\bein{\omega}{\mu}{ab}M_{ab}$
in the basis above and define the invariant tensor to be
 \bea \la{invt}
  \langle M_{AB}M_{CD}M_{EF}\rangle\ = \ \varepsilon_{ABCDEF}\;.
 \eea
Note that, as required, this tensor is \textit{symmetric} in the
sense that it stays invariant under exchange of $M_{AB}$ with
$M_{CD}$, etc. The $SO(4,1)$ covariant field strength tensors in
$F_{\mu\nu}=\frac{1}{2}\bein{{\cal
R}}{\mu\nu}{ab}M_{ab}+\bein{T}{\mu\nu}{a}P_a$ read
 \begin{eqnarray} \label{fieldstr} \begin{split}
  \bein{{\cal R}}{\mu\nu}{ab} &\ =\
  \bein{R}{\mu\nu}{ab}
  +\Lambda(\bein{e}{\mu}{a}\bein{e}{\nu}{b}-\bein{e}{\nu}{a}\bein{e}{\mu}{b})\;,
  \\
  \bein{T}{\mu\nu}{a} &\ =\ \partial_{\mu}\bein{e}{\nu}{a}-\partial_{\nu}\bein{e}{\mu}{a}
  +\bein{\omega}{\mu}{ab}e_{\nu b}-\bein{\omega}{\nu}{ab}e_{\mu
  b}\;, \end{split}
 \eea
containing the Riemann tensor
 \bea
  \bein{R}{\mu\nu}{ab}\ =\ \partial_{\mu}\bein{\omega}{\nu}{ab}
  -\partial_{\nu}\bein{\omega}{\mu}{ab}
  +\omega^{\hspace{0.5em}a}_{\mu\hspace{0.5em}c}\bein{\omega}{\nu}{cb}
  -\omega^{\hspace{0.5em}a}_{\nu\hspace{0.5em}c}\bein{\omega}{\mu}{cb}\;,
 \eea
and the torsion tensor. The resulting Chern-Simons action can be
written as \cite{Chamseddine:1990gk}
 \bea\label{Lovelock}
  \begin{split}
   S\ =\ 3\int_{M_5}\varepsilon_{a_1...a_5} \big( e^{a_1}\wedge R^{a_2a_3}\wedge R^{a_4a_5}
   &+\frac23 e^{a_1}\wedge e^{a_2}\wedge e^{a_3}\wedge R^{a_4a_5}
   \\
   &+\frac15 e^{a_1}\wedge e^{a_2}\wedge e^{a_3}\wedge e^{a_4}\wedge e^{a_5} \big)\;.
  \end{split}
 \eea
We see that the action is the Einstein-Hilbert action with a
cosmological constant (which we have set $\Lambda=1$), extended by
a $D=5$ Lovelock term. To be more precise, it describes a theory
with dynamical torsion. However, it is still consistent with the
field equations to impose vanishing torsion in order to express
the spin connection in terms of the vielbein. This in turn reduces
the dynamical degrees of freedom to those of the metric, for which
the Einstein equations read
 \bea
  R_{\mu\nu}-\frac{1}{2}Rg_{\mu\nu}-3\Lambda g_{\mu\nu}\ = \
  \frac{1}{32\Lambda}(R\hspace{0.2em}R)_{\mu\nu}\;,
 \eea
where we have introduced the abbreviation
$(R\hspace{0.2em}R)_{\mu\nu}=\varepsilon_{\mu}^{\hspace{0.6em}\rho\sigma\lambda\delta}\varepsilon_{\nu\kappa\tau\gamma\phi}
\bein{R}{\rho\sigma}{\kappa\tau}\bein{R}{\lambda\delta}{\gamma\phi}$.

As it stands, (\ref{Lovelock}) seems to be a purely conventional
type of Lovelock gravity, which is usually assumed to propagate
the same number of degrees of freedom as Einstein gravity (five in
$D=5$). However, in this case the topological origin
(\ref{CSgauge}) actually gives rise to a somewhat unconventional
behavior: Expanding (\ref{Lovelock}) around the $AdS_5$ solution
one infers that the quadratic term vanishes identically. In other
words, a propagator around $AdS_5$ does not exist. This can be
most easily understood from the general form of the equations of
motion for the Chern-Simons action (\ref{CSgauge}), which can be
read off from (\ref{CSvar})
 \bea\label{CSeom}
  g_{\cA\cB{\cal C}} F^{\cB}\wedge F^{{\cal C}}\ =\ 0\;,
 \eea
where $g_{\cA\cB{\cal C}}$ denotes the invariant tensor and
$\small \cA$, $\cB$,\ldots, are the adjoint indices for a generic
gauge group, which will be later on specified to $\mhso(4,2)$.
Since for an expansion around $AdS_5$ the curvature tensor in
(\ref{fieldstr}) vanishes in the background, there are no linear
terms in (\ref{CSeom}) and thus no quadratic terms in the action.
However, this should not be interpreted in the sense that the
theory is devoid of local dynamics altogether, as is sometimes
assumed of `topological' actions like (\ref{CSgauge}) in the
literature. Indeed, the propagator around generic backgrounds does
not vanish. Moreover, a careful Hamiltonian analysis of the
dynamical content in \cite{Banados:1995mq,Banados:1996yj} has
shown that, apart from degenerate sectors (like the maximally
symmetric $AdS_5$ background), the theory consistently propagates
a number of degrees of freedom depending on the dimension of the
gauge group. In particular, the Lovelock-type gravity theory above
has the expected five degrees of freedom.\footnote{To be more
precise, this counting applies only in case of vanishing torsion.
Otherwise
 there are additional degrees of freedom \cite{Banados:1996yj}.}
Let us also stress that the degenerate sectors are only a
measure-zero subspace within phase space
\cite{Banados:1995mq,Banados:1996yj}, and that even around such
degenerate backgrounds some degrees of freedom can propagate,
albeit fewer. One example has been given already in
\cite{Chamseddine:1990gk}: It is effectively an $AdS_4$ solution
and reads ($\a,\b=0,1,2,3$)
 \bea
  \bein{\bar{e}}{\mu}{a}=\delta_{\mu}^a\frac{1}{1-\frac14\Lambda
  x^{\alpha}x_{\alpha}}\;, \quad
  \bein{\bar{\omega}}{\mu}{\alpha\beta}=-\frac{\Lambda}{2}
  \frac{\delta_{\mu}^{\alpha}x^{\beta}-\delta_{\mu}^{\beta}x^{\alpha}}
  {1-\frac14\Lambda x^{\alpha}x_{\alpha}}\;, \quad
  \bein{\bar{e}}{4}{4} = \text{const.}\;,
 \eea
which has vanishing torsion, $\bar{T}^a=0$, and satisfies
 \bea\label{4Dsol}
  \bar{R}^{\alpha\beta}+\Lambda\bar{e}^{\alpha}\wedge\bar{e}^{\beta}
  \ =\ 0\;, \qquad
  \bar{R}^{\alpha 4}+\Lambda\bar{e}^{\alpha}\wedge\bar{e}^4\ \neq \ 0\;.
 \eea
By expanding around this solution, it has been shown that it
propagates in particular a four-dimensional graviton
\cite{Chamseddine:1990gk}.

\scss{Invariant Tensor of the Higher-Spin Algebra}
\la{sec:invtens} In order to construct the Chern-Simons action
\eq{CSgauge} based on $\mhso(D-1,2)$, which extends standard
Chern-Simons gravity, we have to find a completely symmetric
tensor of degree $D-2$, which is invariant under the adjoint
action of the HS algebra $\mhso(D-1,2)$, and which reduces to the
standard invariant \eq{invt} for the $AdS$-subalgebra
$\mso(D-1,2)$. Below we will propose a formula for the invariant
tensor. However, while the vector oscillator formulation described
in sec.~\ref{sec:ahse} was required in order to establish
existence and consistency of the HS algebra, it turns out not to
be sufficient for the definition of a symmetric invariants to
$\mhso(D-1,2)$. Instead, we will introduce a new star product,
known as the BCH (Baker-Campell-Hausdorff) star product or the
Gutt star product \cite{gutt:1983,Madore:2000en,Jurco:2000ja}.

Let us first briefly comment on the reasons why the formulation in
terms of vector oscillators is incapable of reproducing the
symmetric tensor (\ref{invt}). This is simply due to the fact that
the oscillators automatically eliminate the totally antisymmetric
part in the star product, $M_{[AB}\ystar M_{CD}\ystar M_{EF]}=0$,
since it involves an antisymmetrization over more than two
$\msp(2)$ indices. On the other hand, this exclusion guaranteed
the appearance of generators entirely being in definite
$(s-1,s-1)$ Young tableaux, or in other words, eliminated the
ideals spanned by generators not in these Young tableaux. Here in
contrast, by requiring an invariant tensor generalizing
(\ref{invt}), we are, roughly speaking, assigning a non-zero value
to certain parts in the ideal ${\cal I}$. Put differently, instead
of using the invariance of the ideal, $[\mhso(D-1,2),{\cal
I}]\subset{\cal I}$, to set it to zero, we set it to constants,
reducing in particular to (\ref{invt}).

To start with, we have to define a non-commutative star product
directly in terms of the $M_{AB}$ (here viewed as commuting
coordinates), whose star commutator then yields the required
$\mso(D-1,2)$ algebra. This is the BCH star product, which is
given by \bea \la{moyalLie} F(M)\star
G(M)&=&\exp\Big(M_{AB}\L^{AB}(\del_N},{\del_{N'})\Big)F(N)G(N')\Big|_{N=M,
N'=M }\ , \eea where $\del_N$ is a short-hand notation for
$\del/\del N_{AB}$ and where $\L^{AB}=-\L^{BA}$ is defined through
the relation \bea \exp
Q\exp{Q'}&=&\exp\big(Q+Q'+\L^{AB}(Q,Q')M_{AB}\big)\ , \eea with
$Q=Q^{AB}M_{AB}$ and $Q'=Q'^{AB}M_{AB}$ for some anti-symmetric
tensors $Q^{AB}$ and $Q'^{AB}$. It defines an associative product
on the enveloping algebra \cite{gutt:1983}. By using the BCH
formula \bea \exp Q\exp{Q'} =
\exp\Big(Q+Q'+\frac12[Q,Q']+\frac{1}{12}\big([Q,[Q,Q']]+[Q',[Q',Q]]\big)+\cdots\Big)\
, \eea we find the first few terms in the expansion to be \bea \nn
\L^{AB}&=&\frac12 f_{CD,EF}{}^{AB}Q^{CD}Q'^{EF}-\frac{1}{12}
f_{CD,EF}{}^{GH}f_{GH,IJ}{}^{AB}Q^{CD}Q'^{EF}(Q^{IJ}-Q'^{IJ})+\cdots
\\ \la{fftex} &=&
2(QQ')^{[AB]}-\frac23[Q,Q']^{[A|C|}(Q_C{}^{B]}-Q'_C{}^{B]})+\cdots\
, \eea where $(QQ')^{AB}=Q^{AC}Q'_C{}^B$,
$[Q,Q']^{AB}=(QQ')^{AB}-(Q'Q)^{AB}$ and where $f_{AB,CD}{}^{EF}$
are the structure constants defined in \eq{structco}. The first
terms in the product \eq{moyalLie} consequently becomes \bea \nn
F(M)\star G(M)&=&F(M)G(M)+2M_{AB}\del^{AC}F\del_C{}^BG
\\ \la{eopt} &&+\frac23M_{AB}\Big(\del^{AC}\del^{BD}F\del_{CD}G-(\del^2)^{AC}F\del^B{}_CG\\
&& \qquad \qquad
+\del^{AC}\del^{BD}G\del_{CD}F-(\del^2)^{AC}G\del^B{}_CF\Big)+\cdots\
. \nn \eea

The definition of the HS generators in \eq{defgen} extends
immediately. However, whereas the realization in terms of the
vector oscillator automatically imposes the Young tableau
symmetries $(s-1,s-1)$, here we need to Young project
explicitly.\footnote{Note that under the projector $\mathbb
P_{(s-1,s-1)}$ the use of the star product or the point-wise
(`classical') product is immaterial. For instance, for spin $3$ we
have $T_{AB,CD}=\mathbb P_{(2,2)}\big(M_{AC}\star
M_{BD}\big)=\mathbb P_{(2,2)}\big(M_{AC}M_{BD}\big)$.} Hence, all
elements of the enveloping algebra which belong to other Young
tableaux are modded out.

The star-products between a spin-$2$ generator and spin-$s$
generator $T_{C(s-1),D(s-1)}$ read \bea M_{AB}\star M_{CD}& = &
M_{AB}M_{CD}-2
\eta\undersym{{}_{C\phantom{\langle}\hspace{-.1cm}[A}M_{B]}\hspace{.35cm}}{}_{\hspace{-.35cm}D}\ ,  \\
\nn M_{AB}\star T_{C(s-1),D(s-1)}& = & M_{AB} T_{C(s-1),D(s-1)}\nn
\\ \la{s2ss} &&-2(s-1)\mathbb
P_{(s-1,s-1)}\big(\eta\undersym{{}_{AC_{s-1}}T_{\phantom{|}B\phantom{|}}}\hspace{-.1cm}{}_{C(s-2),D(s-1)}\big)
\\ && \nn +{\rm ~double~contractions} \ , \eea
where $\mathbb P_{(s-1,s-1)}$ is a Young projector. The
commutation relations in \eq{so42alg} and \eq{spintwos} follow
readily by defining the bracket $[U,V]_\star=U\star V-V\star U$,
since we know that
$[M_{AB},F(M)]_\star=4M_{C[A}\del^C{}_{B]}F(M)$; see Eq.~\eq{abcx}
in appendix \ref{sec:appB}.

Let us now proceed with defining the symmetric invariant tensors
of the HS algebra. Given an element $F(M)$ of the enveloping
algebra $\cU[\mso(D-1,2)]$, we define the operation `$\tr$' given
by evaluation at $M_{AB}=0$ \bea \la{strace} &&\tr \big(F(M)\big)\
:= \ F(0) \ . \eea However, although the analogue of this
operation for the vector oscillator described in
sec.~\ref{sec:ahse} constitutes a proper (super) trace
\cite{Vasiliev:1986qx,Engquist:2005yt,Pinczon}, it is easy to
realize that the bilinear $\tr(F(M)\star G(M))$ vanishes
identically in our case (see also the comments in footnote
\ref{ftosc} below). To obtain a sensible non-zero trace, we need
to insert $GL(D+1)$-invariant differential operators into the
trace \eq{strace} {\it cf.}~the results in
Ref.~\cite{bieliavsky-2002}. A natural $GL(D+1)$-invariant
differential operator is constructed out of $n=(D+1)/2$
derivatives contracted with the totally anti-symmetric tensor
$\epsilon_{A_1\cdots A_{D+1}}$. We propose the following sequence
of traces `$\Tr_k$', for $k=1,2,3,\ldots$ \bea \la{grcycl}
&&\Tr_k\big(F(M)\big)\ = \ \tr\big(\D^k[F(M)]\big) \ , \\
\la{Ddefn} &&\D \ =\ \e_{A_1\cdots A_nB_1\cdots B_n}
\frac{\del}{\del M_{A_1B_1}}\cdots \frac{\del}{\del M_{A_nB_n}}\ ,
\eea with `$\tr$' as in \eq{strace}.\footnote{\la{ftosc} The
oscillator algebra based on \eq{moyalvector} admits a natural
graded (super) trace $\tr_y f(y)=f(0)$, such that
$\tr_y\big(f(y)\ystar g(y)\big)=\tr_y\big(g(-y)\ystar f(y)\big)$
\cite{Vasiliev:1986qx,Engquist:2005yt,Pinczon}. Using this trace
we can construct the anti-symmetric invariants of the
$\mhso(D-1,2)$ algebra. However, for the reasons explained above,
even `dressing' this trace with derivative operators analogous to
\eq{Ddefn}, cannot give rise to a non-vanishing symmetric
combination.} These traces are cyclic \bea \la{cyclicFG}
&&\Tr_k\big(F\star G\big)\ = \ \Tr_k\big(G\star F\big)\ , \eea for
generic elements $F(M)$ and $G(M)$ of the enveloping algebra,
which will be proven in appendix \ref{sec:appB}.

We now define the symmetric trilinear for three generators
$\eq{defgen}$ of the HS algebra $\mhso(4,2)$ of spins $s$, $s'$
and $s''$ to be \bea \la{invte} &&\big\langle
T_s,T_{s'},T_{s''}\big\rangle\ := \
\sum_{k=1}^\infty\a_k\Tr_k\big(\{T_s,T_{s'}\}_\star \star
T_{s''}\big)\ , \eea where $\{T_s,T_{s'}\}_\star=T_s\star
T_{s'}+T_{s'}\star T_s$ and $\a_k$ are arbitrary coefficients.
This definition generalizes directly to an $n$-form for $D=2n-1$.
The total symmetry of \eq{invte} follows from (\ref{cyclicFG}) and
the associativity of the BCH star product.

At this stage we have to note that, strictly speaking, the
cyclicity \eq{cyclicFG} is not sufficient to prove invariance of
\eq{invte}, since the commutator with respect to the BCH star
product potentially contains ideal terms. However, for the
linearization in case of a spin-3 field to be analyzed below, one
can check explicitly that the tensor is invariant to that order.
So we expect \eq{invte} to be invariant under the adjoint action
of the full $\mhso(4,2)$, which, furthermore, might fix the free
coefficients $\alpha_k$.

The definition \eq{invte} will reproduce the symmetric spin-$2$
trilinear \eq{invt} provided we choose the first coefficient to be
$\a_1=1/12$. Further, it follows that the spin-$2$, spin-$2$,
spin-$s$ invariant vanishes for $s>2$ once the symmetries imposed
by the Young projector of the spin-$s$ generator are taken into
account, \bea\label{223tr} \langle
M_{AB},M_{CD},T_{E(s-1),F(s-1)}\rangle\ = \  0\ . \eea This
relation guarantees that the equations of motion for the HS fields
(see (\ref{CSeom})) will not contain a term depending only on the
space-time curvature, which in turn implies that the spin-2 field
does not provide a source for the HS fields. Put differently, it
is consistent with the field equations to set the HS fields to
zero.

Only the first trace $\Tr_1$ enters the linearized spin-$3$
analysis which we will focus on below. The relevant invariant in
$D=5$ is given by \bea \la{relinv} && \langle
M_{AB},T_{CD,EF},T_{GH,IJ}\rangle\ = \ -2\mathbb P_{(2,2)}\mathbb
P'_{(2,2)}\big( \e_{ABCEGI}~ \eta_{DH}\eta_{FJ}\big) \ , \eea
where the projectors impose the symmetries of the two spin-$3$
generators. We note that up to an overall constant, the invariant
\eq{relinv} is the only possible term which is consistent with the
imposed Young symmetries.

Up to now we established the existence of a HS Lie algebra and an
associated symmetric invariant tensor. This in turn is sufficient
to define a consistent HS Chern-Simons action, which in, say,
$D=5$ is given by
 \bea\label{CS}\la{cs}
  S=\int_{M_5}\Big{\langle} {\cal W}\wedge d{\cal
  W}\wedge
  d{\cal W}+\frac32d{\cal W}\wedge {\cal W}\wedge {\cal W}\wedge {\cal W}
  +\frac35{\cal W}\wedge
  {\cal W}\wedge {\cal W}\wedge {\cal W}\wedge {\cal W} \Big{\rangle}\; .
 \eea
Here ${\cal W}$ denotes the gauge field taking values in
${\mhso(4,2)}$. It contains by construction the Lovelock gravity
discussed in sec.~\ref{sec:love}, corresponding to the subalgebra
$\mso(4,2)$. Note that all the complexity of this theory is
encoded in the infinite-dimensional Lie algebra ${\mhso(4,2)}$ and
the symmetric tensor. By virtue of the consistency of
${\mhso(4,2)}$ and the existence of the tri-linear tensor, this
action is by construction invariant under an exact HS symmetry at
the full non-linear level, i.e.~it satisfies requirement (i) in
the introduction. However, due to the fact that the Lie brackets
of ${\mhso(4,2)}$ are not known explicitly, at this stage the
action (\ref{CS}) cannot be rewritten in a closed form in terms of
the physical HS fields. Fortunately a linearized analysis can be
performed, and in the next section we will show that one recovers
indeed the correct free field limit, thus proving that (\ref{CS})
satisfies also condition (ii).

\scs{Dynamical Analysis}\la{sec:DA} In this section we will
discuss some aspects of the dynamical content of the constructed
HS theory. As it stands, the HS action (\ref{cs}) describes a
theory with propagating gravitational torsion, so we expect also
the HS torsions (which will be defined below) to propagate. Since
the dynamics of these kind of theories is much less understood, we
take here a pragmatic point of view, i.e.~we impose vanishing
torsion, which is compatible with the equations of motion though
it is not enforced by them. For simplicity our focus will be on
the first non-trivial case, {\it viz.}~spin-$3$, which we believe
exhibits generic features present for arbitrary spin.

\scss{Linearization and Constraints for Spin-$3$}\label{constrsec}
We first note that, as in the purely gravitational case, an
expansion around $AdS_5$ does not give rise to a non-trivial
propagator. This can be seen by inspecting the equations of motion
(\ref{CSeom}). Up to first order they are of the form
 \bea\label{CSHSeq}
  g_{\cal ABC}{\cal R}_{\rm AdS}^{\cal B}\wedge R_{\rm HS}^{\cal C}\ = \ 0\;,
 \eea
where $R_{\rm HS}$ denotes the linearized HS contribution. As the
$AdS$-covariant field strength vanishes in the $AdS$ background,
${\cal R}_{\rm AdS}=0$, the equations are identically satisfied at
the first order and do not lead to any perturbative dynamics.

Instead we will first keep the discussion generic and later focus
on an expansion around the $AdS_4\times S^1$ solution discussed in
sec.~\ref{sec:love}. For this we have to know the HS algebra
explicitly. Fortunately, for an expansion around a given
background geometry, only the commutators between spin-2 and
spin-$s$ generator enter, while the mutual interactions between
the different HS fields are not relevant. The spin-3 generator is
given by $T_{AB,CD}$, corresponding to the Young tableau
$\tiny{\yng(2,2)}$\ , and it closes according to (\ref{spintwos})
with the spin-2 generator as\footnote{In the sequel we will drop
the $\star$ subscript on the commutators.}
 \begin{eqnarray}
  [M_{AB},T_{CD,EF}]&=& \nonumber -2\Big(\eta\undersym{{}_{A\phantom{\langle}\hspace{-.1cm}C}T_B\hspace{.1cm}}{}_{\hspace{-.1cm}D,EF}
  +\eta\undersym{{}_{A\phantom{\langle}\hspace{-.1cm}D}T_{CB\hspace{.1cm}}}{}_{\hspace{-.1cm},EF}
  +\eta\undersym{{}_{AE}T_{CD,}{}_B\hspace{.1cm}}{}_{\hspace{-.1cm}F}
  +\eta\undersym{{}_{AF}T_{CD,E}{}_B\hspace{.1cm}}\Big)\;,   \\
  &=&-8\eta\undersym{{}_{A\langle C}T_{|B|}}{}_{D,EF\rangle}\ .
 \end{eqnarray}
Here curly brackets denote $(2,2)$ Young projection, while in the
following they also indicate symmetrization according to the Hook
tableau, etc. (see appendix \ref{A}). In a $GL(5)$ covariant
basis, the spin-3 generators are given by $T_{ab}=T_{ab,66}$,
$T_{ab,c}=T_{ab,c6}$ and $T_{ab,cd}$, and their algebra reads
 \begin{eqnarray}\label{spin23}\begin{split}
  [M_{ab},T_{cd}] &\ =\ -4\eta\undersym{{}_{a\langle c}T_{|b|\hspace{.1cm}}}{}_{\hspace{-.1cm}d\rangle}\;, \qquad
  [M_{ab},T_{cd,e}]\ = \ -4\eta\undersym{{}_{a\langle c}T_{|b|\hspace{.1cm}}}{}_{\hspace{-.1cm}d,e\rangle}+4\eta\undersym{{}_{a\langle c}T_{de\rangle,{}\hspace{.2cm}\phantom{b}}}\hspace{-.3cm}{}_b\;, \\
  [M_{ab},T_{cd,ef}] &\ =\ -8\eta\undersym{{}_{a\langle c}T_{|b|\hspace{.1cm}}}{}_{\hspace{-.1cm}d,ef\rangle}\;, \qquad
  [P_a,T_{bc}]\ = \-2T_{bc,a}\;, \\
  [P_a,T_{bc,d}] &\ =\ 3\eta_{a\langle b}T_{cd\rangle}-T_{ad,bc}\;, \qquad
  [P_a,T_{bc,de}]\ = \ 8\eta_{a\langle b}T_{cd,e\rangle}\;. \end{split}
 \end{eqnarray}
Here we take the brackets $[T,T]$ to be vanishing, even though in
the full HS algebra they close into spin-4 generator. However, in
the linearization these spin-4 fields decouple, and, indeed, this
truncation defines a consistent Lie algebra.

Next we linearize the HS gauge field as\footnote{The unit-strength
normalizations follow from the Hook length formula
\cite{Fuchs:1997jv}.}
 \bea\label{gaugeexp}
  {\cal W}_{\mu}\ = \ \bein{\bar{e}}{\mu}{a}P_a +
  \frac{1}{2}\bein{\bar{\omega}}{\mu}{ab}M_{ab} +
  \kappa\big(\frac{1}{2}\bein{e}{\mu}{ab}T_{ab}+\frac{1}{3}\bein{\omega}{\mu}{ab,c}
  T_{ab,c}+\frac{1}{12}\bein{\omega}{\mu}{ab,cd}T_{ab,cd}\big)\;,
 \eea
where $\bein{\bar{e}}{\mu}{a}$ and $\bein{\bar{\omega}}{\mu}{ab}$
are vielbein and spin connection of the background geometry.
Moreover, we consistently omitted contributions from all fields
with spin $s>3$. $\bein{e}{\mu}{ab}$ will later be identified with
the spin-3 field, while $\bein{\omega}{\mu}{ab,c}$ and
$\bein{\omega}{\mu}{ab,cd}$ are auxiliary fields that have to be
eliminated by means of constraints. It will turn out that these
constraints are analogous to the torsion constraint of general
relativity. As the torsion tensor appears as part of the field
strength in (\ref{fieldstr}), we will determine the required
constraints in the HS case by computing the non-abelian field
strength based on the algebra (\ref{spin23}). We find
 \begin{eqnarray}
  {\cal F}_{\mu\nu} &=& \partial_{\mu}{\cal W}_{\nu} -\partial_{\nu}{\cal W}_{\mu}
  +[{\cal W}_{\mu},{\cal W}_{\nu}] \\ \nonumber
  &=& \bein{\bar{T}}{\mu\nu}{a}P_a +
  \frac{1}{2}\bein{\bar{{\cal R}}}{\mu\nu}{ab}M_{ab}+\kappa
  \big(\frac{1}{2}\bein{T}{\mu\nu}{ab}T_{ab} +
  \frac{1}{3}\bein{T}{\mu\nu}{ab,c}T_{ab,c}
  +\frac{1}{12}\bein{R}{\mu\nu}{ab,cd}T_{ab,cd}\big) +
  \cO(\kappa^2)\;.
 \end{eqnarray}
Here $\bein{\bar{T}}{\mu\nu}{a}$ denotes the background torsion,
which we assume to vanish, while $\bein{\bar{{\cal
R}}}{\mu\nu}{ab}$ is the $AdS$-covariant background curvature
tensor. The linearized HS field strengths read
 \begin{eqnarray} \begin{split}
  \bein{T}{\mu\nu}{ab} &\ =\ \bar{D}_{\mu}\bein{e}{\nu}{ab}-
  \bar{D}_{\nu}\bein{e}{\mu}{ab}
  +\bein{\omega}{\mu}{ab,c}\bar{e}_{\nu c}-
  \bein{\omega}{\nu}{ab,c}\bar{e}_{\mu c} \;,        \\ \label{strength}
  \bein{T}{\mu\nu}{ab,c} & \ = \
  \bar{D}_{\mu}\bein{\omega}{\nu}{ab,c}-\bar{D}_{\nu}\bein{\omega}{\mu}{ab,c}
  +\bein{\omega}{\mu}{ab,cd}\bar{e}_{\nu d}-\bein{\omega}{\nu}{ab,cd}\bar{e}_{\mu
  d}\\
  &\qquad +3\bein{e}{\mu}{\langle ab}\bein{\bar{e}}{\nu}{c\rangle}-
  3\bein{e}{\nu}{\langle ab}\bein{\bar{e}}{\mu}{c\rangle}    \;, \\
  \bein{R}{\mu\nu}{ab,cd} &\ =\
  \bar{D}_{\mu}\bein{\omega}{\nu}{ab,cd}-\bar{D}_{\nu}\bein{\omega}{\mu}{ab,cd}
  +4\bein{\omega}{\mu}{\langle ab,c}\bein{\bar{e}}{\nu}{d\rangle}
  -4\bein{\omega}{\nu}{\langle ab,c}\bein{\bar{e}}{\mu}{d\rangle}
  \;,
 \end{split}
 \end{eqnarray}
where $\bar{D}_{\mu}$ denotes the background Lorentz covariant
derivative, which read on the different fields
 \begin{eqnarray} \nonumber
  \bar{D}_{\mu}\bein{e}{\nu}{ab}&=&\partial_{\mu}\bein{e}{\nu}{ab}
  +2\spinleg{\bar{\omega}}{\mu}{\langle a}{c}
  \bein{e}{\nu}{|c|b\rangle}\;,\\
  \bar{D}_{\mu}\bein{\omega}{\nu}{ab,c} &=& \partial_{\mu}\bein{\omega}{\nu}{ab,c}
  +2\hspace{0.1em}\spinleg{\bar{\omega}}{\mu}{\langle a}{d}\bein{\omega}{\nu}{|d|b,c\rangle}
  -2\hspace{0.1em}\spinleg{\bar{\omega}}{\mu}{\langle a}{d}\bein{\omega}{\nu}{bc\rangle,d}
  \;, \\ \nonumber
  \bar{D}_{\mu}\bein{\omega}{\nu}{ab,cd} &=& \partial_{\mu}\bein{\omega}{\nu}{ab,cd}
  +4\spinleg{\bar{\omega}}{\mu}{\langle a}{e}\bein{\omega}{\nu}{|e|b,cd\rangle} \;.
 \end{eqnarray}

Before we turn to the constraints let us discuss the spin-3
symmetries, under which the field strengths above stay invariant.
Under a non-abelian gauge transformation $\delta {\cal W}_{\mu} =
D_{\mu}\epsilon =
\partial_{\mu}\epsilon+[{\cal W}_{\mu},\epsilon]$, with Lie algebra valued
transformation parameter $\epsilon$ given in the spin-3 case by
 \bea
  \epsilon\ = \ \xi^a P_a +\frac{1}{2}\Lambda^{ab}M_{ab}
  +\kappa\big(
  \frac{1}{2}\epsilon^{ab}T_{ab}+\frac{1}{3}\epsilon^{ab,c}T_{ab,c}
  +\frac{1}{12}\epsilon^{ab,cd}T_{ab,cd}\big)\;,
 \eea
we find the following variations (ignoring background
diffeomorphisms and Lorentz transformations)
 \begin{eqnarray} \nonumber
  \delta_{\epsilon}\bein{e}{\mu}{ab} &=&
  \bar{D}_{\mu}\epsilon^{ab}-\epsilon^{ab,c}\bar{e}_{\mu c}\;, \\
  \label{hssym}
  \delta_{\epsilon}\bein{\omega}{\mu}{ab,c} &=&
  \bar{D}_{\mu}\epsilon^{ab,c}-3\epsilon^{\langle ab}\bein{\bar{e}}{\mu}{c\rangle}
  -\epsilon^{ab,cd}\bar{e}_{\mu d}\;, \\ \nonumber
  \delta_{\epsilon}\bein{\omega}{\mu}{ab,cd} &=& \bar{D}_{\mu}\epsilon^{ab,cd}
  -4\epsilon^{\langle ab,c}\bein{\bar{e}}{\mu}{d\rangle}\;.
 \end{eqnarray}
Note that the gauge transformations with parameter
$\epsilon^{ab,c}$ and $\epsilon^{ab,cd}$ corresponding to the
auxiliary fields act as St\"uckelberg shift symmetries.

Next we are going to discuss the constraints. We will see that
imposing the conditions\footnote{Note that the first constraint
allows to identify the background diffeomorphisms with the gauge
transformations generated by $\xi^a$ in the sense that the latter
read on $\bein{e}{\mu}{ab}$, up to local Lorentz and St\"uckelberg
transformations,
$\delta_{\xi}\bein{e}{\mu}{ab}=\pounds_{\xi}\bein{e}{\mu}{ab}$,
where $\pounds_{\xi}$ denotes the Lie derivative with respect to
the vector field
$\xi^{\mu}=\bar{e}^{\mu}_{\hspace{0.3em}a}\xi^a$.}
 \bea\label{constr}
  \bein{T}{\mu\nu}{ab}\ = \ 0\;, \qquad
  \bein{T}{\mu\nu}{ab,c}\ = \ 0\;,
 \eea
allows to express $\bein{\omega}{\mu}{ab,c}$ in terms of the
physical spin-3 field $\bein{e}{\mu}{ab}$ and its first derivative
and $\bein{\omega}{\mu}{ab,cd}$ in terms of
$\bein{\omega}{\mu}{ab,c}$ and its first derivatives. In turn,
$\bein{\omega}{\mu}{ab,cd}$ is a function of $\bein{e}{\mu}{ab}$
and its first and second derivatives. The latter can be inserted
into the third of the equations (\ref{strength}), which then
yields the HS generalization of the Riemannian curvature tensor.
Therefore the spin-3 curvature tensor will be of third order in
the derivatives of the spin-3 field. This procedure can be
generalized to arbitrary spin-$s$ fields, whose curvature tensor
will thus contain the $s$-th derivative of the physical spin-$s$
field. (For trace-less tensors in $D=4$ spinorial form this
analysis has been done in \cite{Vasiliev:1986td}, while a
cohomological analysis in $D$ dimensions can be found in
\cite{Lopatin:1987hz,Bekaert:2005vh}.) This corresponds to the
hierarchy of de\hspace{-0.3em} Wit--Freedman connections found in
the metric-like formulation \cite{deWit:1979pe}. Since the
equations of motion will necessarily impose conditions on the HS
Riemann tensor, this implies that the field equations are in the
linearization already of higher derivative order. So at first
sight we seem to have little chance to recover the required
$2^{\rm nd}$ order Fr\o nsdal equations. However, in flat space it
has been shown that the Riemann tensor is a curl (`Damour-Deser
identity' \cite{Damour:1987vm}) and that it can therefore be
locally integrated, giving rise to the Fr\o nsdal equations in the
so-called compensator formulation
\cite{Sagnotti:2005ns,Bekaert:2003az}. Here we will prove that
this generalizes to $AdS$.

Let us now turn to the constraints. From the first of the
equations (\ref{constr}) we conclude
 \bea\label{anhol}
  \omega^{d\hspace{0.1em}bc,a}-\omega^{a\hspace{0.1em}bc,d}\ =\
  \Omega_1^{\hspace{0.2em}ad,bc}\;,
 \eea
where the curved index on $\bein{\omega}{\mu}{ab,c}$ has been
converted into a flat index by means of the background vielbein,
and we have introduced a HS generalization of the coefficients of
anholonomity,
 \bea
  \Omega_1^{\hspace{0.2em}ab,cd}\ = \ \bar{e}^{\mu a}\bar{e}^{\nu b}
  \big(\bar{D}_{\mu}\bein{e}{\nu}{cd}-\bar{D}_{\nu}\bein{e}{\mu}{cd}\big)\;.
 \eea
By permuting the indices in (\ref{anhol}), one finds the
expression
 \bea\label{torsol}
  \bein{\omega}{\mu}{bc,d}\ = \ \frac{1}{2}
  \bar{e}_{\mu a}\big(\Omega_1^{\hspace{0.2em}a(b,c)d}-\Omega_1^{\hspace{0.2em}ad,bc}
  +\Omega_1^{\hspace{0.2em}d(b,c)a}\big)+\bein{\xi}{\mu}{bc,d}\;,
 \eea
where
 \bea\label{xi}
  \bein{\xi}{\mu}{bc,d}\ = \ \frac{1}{4}\bar{e}_{\mu
  a}\big(\omega^{abc,d}+\omega^{bda,c}+\omega^{cda,b}+\omega^{dbc,a}\big)\;.
 \eea
To understand the significance of $\bein{\xi}{\mu}{ab,c}$, we
first note that a priori (\ref{torsol}) lives in the Young
tableaux
 \bea
  {\small  \yng(1)} \hspace{0.5em} \otimes\hspace{0.5em}  {\small \yng(2,1) }
  \hspace{0.5em}=\hspace{0.5em}
  {\small \yng(2,2)}\hspace{0.5em}\oplus\hspace{0.5em} {\small
  \yng(2,1,1)}\hspace{0.5em}\oplus\hspace{0.5em} {\small
  \yng(3,1)}\;\;.
 \eea
It follows from (\ref{xi}) that $\xi$ is in the window tableau,
i.e.~$(1-\mathbb{P}_{(2,2)})\xi = 0$. In the following we will
have to treat $\xi$ as an independent field. One can easily check
that (\ref{torsol}) solves (\ref{anhol}) for arbitrary $\xi$, by
using the window property of the latter. In fact, we will see that
the inclusion of this auxiliary field is necessary in order for
the composite connection $\bein{\omega}{\mu}{bc,d}(e,\xi)$ to
reproduce the correct transformation behavior in (\ref{hssym}).

From now on we will specify the geometry to $AdS$, since this is
the case we are interested in later on.\footnote{Note, however,
that the analysis performed here holds in an arbitrary dimension,
i.e.~it applies in particular to $AdS_5$ as well as the $AdS_4$
geometry we will discuss below.} Specifically, the
background-covariant derivative $\bar{D}_{\mu}$ reduces to the
$AdS$-covariant derivative $\nabla_{\mu}$, characterized by
 \bea\label{adscomm}
  [\nabla_{\mu},\nabla_{\nu}]V_{\rho}\ =\
  \frac{1}{L^2}(g_{\nu\rho}V_{\mu}-g_{\mu\rho}V_{\nu})\;,
 \eea
with the $AdS$ metric $g_{\mu\nu}$ of radius $L$, which in our
conventions is $L=1$. Applying the first equation of (\ref{hssym})
to (\ref{torsol}), one can verify by use of (\ref{adscomm}) that
$\bein{\omega}{\mu}{ab,c}(e,\xi)$ transforms exactly as required
by the second equation of (\ref{hssym}), if one defines
 \bea\label{xivar}
  \delta_{\epsilon}\xi_{\mu\hspace{0.3em}\nu\rho,\sigma}\ = \ \nabla_{\langle \mu}
  \epsilon_{\nu\rho,\sigma\rangle}-3\epsilon_{\langle \nu\rho}g_{\sigma\mu\rangle}
  -\epsilon_{\nu\rho,\sigma\mu}\;.
 \eea
In particular one sees that this transformation rule is consistent
with the window symmetry of $\bein{\xi}{\mu}{ab,c}$.

The second torsion constraint in (\ref{constr}) can now be solved
in a similar fashion. For our purposes it will, however, be
sufficient to perform this analysis in a gauged-fixed formulation
(for $AdS$ backgrounds). This will effectively reduce the field
content to the completely symmetry spin-3 field, given in a
metric-like formulation by
 \bea\label{spin3}
  h_{\mu\nu\rho}\ :=\  \bein{\bar{e}}{(\mu}{a}\bein{\bar{e}}{\nu}{b}
  e_{\rho )ab}\;.
 \eea
Specifically we use the St\"uckelberg shift symmetry in
(\ref{hssym}) parametrized by $\epsilon^{ab,c}$ to gauge the
hooked part $\tiny{\yng(2,1)}$ of $e_{\mu ab}$ to zero (see
(\ref{hooktens}) in app.~A). However, this gauge fixing will be
violated by a generic spin-3 transformation, and so one has to add
a compensating shift transformation with parameter
$\epsilon_{ab,c}= \nabla_{\langle c}\epsilon_{ab\rangle}$. Under
this residual gauge symmetry only the completely symmetric part of
$e_{\mu\nu\rho}$ transforms, namely as
 \bea
  \delta_{\epsilon}h_{\mu\nu\rho}\ = \
  \nabla_{(\mu}\epsilon_{\nu\rho )}\;,
 \eea
as required in the free limit (see sec.~\ref{free}). Furthermore,
from (\ref{xivar}) we infer, that also $\bein{\xi}{\mu}{ab,c}$ is
subject to a St\"uckelberg shift symmetry with transformation
parameter in $\tiny{\yng(2,2)}$. Therefore it can be gauged away
completely, which in turn requires a compensating transformation
with
 \bea
  \epsilon_{\nu\rho,\mu\sigma}\ = \
  \nabla_{\langle\mu}\nabla_{\sigma}\epsilon_{\nu\rho\rangle}
  -3\epsilon_{\langle\nu\rho}g_{\sigma\mu\rangle}\;.
 \eea
In total, after gauge-fixing the spin-3 connections will depend
only on the completely symmetric part of $e_{\mu ab}$.

To solve the second torsion constraint we derive from
(\ref{strength}) for $\bein{\omega}{\nu}{ab,cd}$ in flat indices:
 \bea\label{tor2}
  \omega^{a\hspace{0.2em}bc,de}-\omega^{e\hspace{0.2em}bc,da}
  \ =\ \Omega_2^{\hspace{0,2em}ea\hspace{0.2em}bc,d}\;,
 \eea
where
 \bea\label{Omega2}
  \Omega_2^{\hspace{0,2em}ab\hspace{0.2em}cd,e}\ = \
  \bar{e}^{a\mu}\bar{e}^{b\nu}\big(\bar{D}_{\mu}\bein{\omega}{\nu}{cd,e}
  +3\bein{e}{\mu}{\langle cd}\bein{\bar{e}}{\nu}{e\rangle}-(\mu\leftrightarrow
  \nu) \big)\;.
 \eea
We find the solution
 \bea\label{torsol2}
  \bein{\omega}{\mu}{ab,cd}\ = \ -\frac{1}{2}\bar{e}_{\mu f}
  \big(\Omega_2^{\hspace{0,2em}fa\hspace{0.2em}cd,b}
  +\Omega_2^{\hspace{0,2em}fb\hspace{0.2em}cd,a}
  +\Omega_2^{\hspace{0,2em}fc\hspace{0.2em}ab,d}
  +\Omega_2^{\hspace{0,2em}fd\hspace{0.2em}ab,c}\big)
  \;.
 \eea
To verify that this is a solution it is not sufficient to use the
symmetries of the $\Omega_2^{\hspace{0,2em}ab\hspace{0.2em}cd,e}$,
but instead the explicit expression given by (\ref{Omega2}) and
(\ref{torsol}) together with the $AdS$ relation (\ref{adscomm}) is
required.

\scss{Spin-3 Field Equations} Let us now turn to the equations of
motion. We specify to the $AdS_4$ background discussed in
sec.~\ref{sec:love}. Moreover, we set all components of
$\bein{e}{\mu}{ab}$ which have a leg in the fifth dimension to
zero. In other words, we are not considering the dynamics of
Kaluza-Klein scalars and vectors, etc., in order to simplify the
analysis. Though in the full non-linear theory this would most
likely not be a consistent Kaluza-Klein truncation, in the
linearization this is justified since the different fields
decouple.

Using the explicit form of the invariant tensor in \eq{relinv} we
see that, after imposing the constraint, the only non-trivial part
of the equations of motion (\ref{CSHSeq}) requires the free index
to take values in the hook tableau. Moreover, we have seen in
eq.~(\ref{223tr}) that setting the background spin-3 field to zero
is consistent with its equations of motion, which we implicitly
assumed already in the expansion (\ref{gaugeexp}).

Specifically, by use of \eq{relinv} we have
 \bea
  0\ =\ \mathbb{P}_{{\tiny\yng(2,1)}}\left(\epsilon_{abcde}\bar{R}^{ab}\wedge
  R^{d\hspace{0.8em}e}_{\hspace{0.4em}f,\hspace{0.6em}h}\right)
  =\frac{1}{2}\left(\epsilon_{abcde}\bar{R}^{ab}\wedge
  R^{d\hspace{0.8em}e}_{\hspace{0.4em}f,\hspace{0.6em}h}+
  \epsilon_{abhde}\bar{R}^{ab}\wedge
  R^{d\hspace{0.8em}e}_{\hspace{0.4em}f,\hspace{0.6em}c}\right)\;,
 \eea
where we used in the second equation the projector in $(ch,f)$.
Specifying now to flat $AdS_4$ indices
$a=(\alpha,4)$,\footnote{Indices $\mu$, $\nu$,\ldots, denote
$D$-dimensional space-time indices. We hope it will not source any
confusion that we specify them in this section to curved $AdS_4$
indices.} and using (\ref{4Dsol}) this implies
 \bea
  \epsilon_{\alpha\beta\gamma\delta}\bar{e}^{\alpha}\wedge
  R^{\gamma\hspace{0.8em}\delta}_{\hspace{0.4em}\epsilon,\hspace{0.6em}\zeta}
  +(\zeta \leftrightarrow \beta)\  = \ 0 \;.
 \eea
This yields in components by use of the identity
$\epsilon_{\alpha\beta\gamma\delta}\epsilon^{\mu\nu\rho\sigma}
\bein{\bar{e}}{\mu}{\alpha}=3!\hspace{0.1em}e\hspace{0.1em}
\bar{e}_{[\beta}^{\nu}
\bar{e}_{\gamma}^{\rho}\bar{e}_{\delta]}^{\sigma}$ and after
relabelling the indices
 \bea\label{spin3eq}
  R_{\mu\lambda\hspace{0.8em}\nu,\rho\sigma}^{\hspace{1.2em}\lambda}-
  R_{\mu\lambda\hspace{0.8em}\sigma,\rho\nu}^{\hspace{1.2em}\lambda}+
  R_{\lambda\delta\hspace{0.9em}\nu,\hspace{0.5em}\sigma}
  ^{\hspace{1.2em}\lambda\hspace{0.7em}\delta}g_{\mu\rho}+(\mu\leftrightarrow
  \sigma)\ = \ 0\;.
 \eea
Taking the $\mu,\rho$ trace implies that the double trace of the
Riemann tensor vanishes. We prove in appendix \ref{sec:appC} that
the final equation is equivalent to the condition that any single
trace of the Riemann tensor, i.e.~the spin-3 analogue of the Ricci
tensor, vanishes. It turns out that a convenient choice is the
following:
 \bea\label{RieTr}
  R_{\mu\nu\hspace{0.3em}\alpha\beta},{}^\d{}_\d\
  =\
  0\;.
 \eea

Next we are going to analyze this equation in more detail. By
inserting (\ref{torsol}) into (\ref{torsol2}) and using
(\ref{strength}) one finds the explicit expressions (in curved
indices)
 \bea\label{curl}
  \nabla_{\mu}K_{\nu\hspace{0.2em}\rho\sigma}+g_{\sigma\nu}(\nabla_{\rho}h_{\mu}^{\prime}
  -\nabla\cdot h_{\mu\rho})+g_{\rho\nu}(\nabla_{\sigma}h_{\mu}^{\prime}-\nabla\cdot
  h_{\mu\sigma})-(\mu\leftrightarrow\nu)\ =\ 0\;,
 \eea
where we defined
 \bea\label{K}
  \begin{split}
   K_{\nu\hspace{0.2em}\rho\sigma}\ =\ \square
   h_{\nu\rho\sigma}-\nabla_{\rho}\nabla\cdot h_{\sigma\nu}-\nabla_{\sigma}
   \nabla\cdot h_{\rho\nu}&+\nabla_{( \rho}\nabla_{\sigma)}h_{\nu}^{\prime}\\
   &-(D-3)h_{\nu\rho\sigma}
   -g_{\nu (\rho}h_{\sigma)}^{\prime}-3g_{\rho\sigma}h_{\nu}^{\prime}\;.
  \end{split}
 \eea
Here we left the space-time dimension generic, though in our case
it is $D=4$. As outlined above, we are going to show that these
3$^{\rm rd}$ order differential equations can locally be
integrated to give effectively rise to sensible 2$^{\rm nd}$ order
field equations. We first note that, in contrast to Minkowski
space, on $AdS$ a vanishing curl cannot locally be integrated to a
gradient, since the covariant derivatives do not commute. However,
for $K_{\nu\hspace{0.2em}\rho\sigma}$ symmetric in $\rho,\sigma$ a
condition like
 \bea
  \nabla_{[\mu}K_{\nu]\hspace{0.2em}\rho\sigma}\ =\ -2g_{\rho[\mu}\beta_{\nu]\sigma}
  -2g_{\sigma[\mu}\beta_{\nu]\rho}\;,
 \eea
with symmetric $\beta_{\mu\nu}$, can be solved by
$K_{\mu\hspace{0.2em}\nu\rho}=\nabla_{\mu}\beta_{\nu\rho}$, as
follows after reinsertion from (\ref{adscomm}). Comparing with
(\ref{curl}) we see that the equations of motion have almost this
form, except that the $\beta_{\mu\nu}$ derived like this are not
symmetric. If the latter are symmetrized by hand, additional terms
have to be added to the ansatz for
$K_{\mu\hspace{0.2em}\rho\sigma}$, which in turns implies that it
is no longer a pure gradient. Moreover, an integration constant
has to be carefully taken into account. Altogether one finds that
 \bea\label{locsol}
  K_{\nu\hspace{0.2em}\rho\sigma}\ =\ \nabla_{\nu}\beta_{\rho\sigma}
  +g_{\nu\rho}(h_{\sigma}^{\prime}-\nabla_{\sigma}\alpha)+
  g_{\nu\sigma}(h_{\rho}^{\prime}-\nabla_{\rho}\alpha)
 \eea
solves (\ref{curl}), where
 \bea
  \beta_{\mu\nu}\ = \ \nabla\cdot
  h_{\mu\nu}-\nabla_{\mu}h_{\nu}^{\prime}-\nabla_{\nu}h_{\mu}^{\prime}
  +\nabla_{\mu}\nabla_{\nu}\alpha-2g_{\mu\nu}\alpha\;,
 \eea
and $\alpha$ is the integration constant. Then (\ref{locsol}) can
be rewritten by use of the explicit expression in (\ref{K}) as
 \bea\label{sagnotti}
  {\cal F}_{\mu\nu\rho}^{\rm AdS}\ = \
  \nabla_{(\mu}\nabla_{\nu}\nabla_{\rho)}\alpha-4
  g_{(\mu\nu}\nabla_{\rho)}\alpha\;,
 \eea
where ${\cal F}^{\rm AdS}$ is the $AdS$ Fr\o nsdal operator
defined in sec.~\ref{free}. To understand the significance of
$\alpha$ we note that the equation (\ref{curl}) is by construction
spin-3 invariant. However, by locally integrating, this invariance
would be lost, if not a non-trivial transformation behavior is
assigned to the `integration constant' $\alpha$. In fact,
(\ref{sagnotti}) is only invariant if
 \bea
  \delta_{\epsilon}\alpha\ = \ \epsilon^{\prime}\;.
 \eea
This shift symmetry can now be used to set $\alpha = 0$, such that
(\ref{sagnotti}) reduces to the Fr\o nsdal equation on $AdS$, the
latter being invariant under all trace-less spin-3
transformations. Thus we correctly recovered the required free
spin-$3$ equations. The formulation (\ref{sagnotti}) with its
invariance under \textit{trace-full} transformations and the
appearance of the so-called compensator $\alpha$ in fact coincides
completely with the construction of Francia and Sagnotti
\cite{Francia:2002aa,Francia:2002pt,Francia:2006hp}.

\scs{Conclusions and Outlook} By virtue of the Yang-Mills gauge
invariance of Chern-Simons actions in any odd dimension, the HS
theories constructed in this paper provide a consistent coupling
to gravity in the sense that the free HS symmetry
$\delta_{\epsilon}h_{\mu_1...\mu_s}=\nabla_{(
\mu_1}\epsilon_{\mu_2...\mu_s )}$ gets deformed to an exact
symmetry of the full non-linear theory. In other words, condition
(i) raised in the in the introduction is satisfied by
construction. Moreover, contrary to what is sometimes implicitly
assumed, these `topological' actions do possess propagating
degrees of freedom for $D>3$. By linearizing around the $AdS_4$
solution found in \cite{Chamseddine:1990gk}, we verified
explicitly that this is the case especially in the presence of HS
fields. We recovered the correct free field equations in the first
non-trivial case of a spin-3 field. For this we showed that on the
subsector of vanishing spin-3 torsion the field equations, though
being 3$^{\rm rd}$ order differential equations, can locally be
integrated to 2$^{\rm nd}$ order equations, which in turn coincide
with the Fr\o nsdal equations in the formulation of
\cite{Francia:2002aa,Francia:2002pt}.

We would like to stress that this is in contrast to previous
attempts to construct consistent HS actions
\cite{Fradkin:1986qy,Vasiliev:2001wa}: In order to guarantee free
field equations of 2$^{\rm nd}$ order, they impose the additional
condition that the `extra fields' (in our case the spin-3
connection $\bein{\omega}{\mu}{ab,cd}$), which are generically of
higher derivative order, do not enter the free action. Here we do
not have this freedom, since the action is completely determined
by gauge invariance, i.e.~the extra fields inevitably enter the
free theory. That we get nevertheless the correct Fr\o nsdal
equations, or in other words, that the higher-derivatives are
gauge artefacts that can be eliminated, is due to the curl-like
structure of the HS Riemann tensor, which in flat space is known
as the Damour-Deser identity \cite{Damour:1987vm,Bekaert:2003az}
(see also \cite{Bekaert:2006ix,Bandos:2005mb}). Since we verified
here an analogous behaviour on $AdS$ for spin-3, this pattern will
most likely extend to all HS fields, and therefore requirement
(ii) for consistent HS theories is satisfied.

Let us also stress that in this approach it is very natural, if
not necessary, to start with a HS algebra based on trace-full
generators, since then the appearance of the compensator $\alpha$
in the `integration' leading from (\ref{curl}) to (\ref{sagnotti})
has a very natural interpretation in that it compensates for the
non-invariance of the pure Fr\o nsdal operator under trace-full HS
transformations. Moreover, starting with trace-less generators
would imply in particular that the HS Riemann tensor is already
trace-less and consequently the field equations in the form
(\ref{RieTr}) would be identically satisfied and not lead to any
dynamics. Instead, the dynamics could possibly be encoded, via the
Bianchi identities, in the lower-rank torsion-like tensors, for
which, however, the distinction between constraint equations and
dynamical equations would be less straightforward. (See also the
discussion about the so-called $\sigma^-$-- cohomology in
\cite{Bekaert:2005vh} and references therein.)

Finally we note that, compared to the `unfolded formulation' of HS
theories advertised in the literature so far, the more
conventional action principle presented here has the advantage of
admitting already a class of exact solutions. In fact, by virtue
of the relation (\ref{223tr}) we concluded that any solution of
the purely gravitational theory, as for instance black holes
\cite{Banados:1997df} and pp-waves \cite{Edelstein:2006se}, can be
lifted to an exact solution of the full theory, simply by setting
all HS fields to zero. Accordingly, this theory allows the
analysis of HS dynamics on more complicated backgrounds (and then,
in principle, also of the back reaction of the geometry). This is
in contrast to the unfolded dynamics, for which even in case that
all HS fields vanish, the construction of solutions is a highly
non-trivial problem. Indeed, apart from $AdS$, exact solutions
have been found only recently \cite{Sezgin:2005pv,Didenko:2006zd}.

Many things are left to be done. First, we have analyzed the
dynamical content only in case of trivial HS torsion. However,
viewed as a 1$^{\rm st}$ order formulation, the theory does not
imply vanishing torsion (though the latter provides a particular
solution). So either one imposes the torsion constraints by hand,
in order to express the de\hspace{-0.3em} Wit--Freedman
connections in terms of the physical HS fields, or one treats the
torsions as carrying additional degrees of freedom. In the former
case it is not clear that this is consistent with the HS gauge
symmetry: Although we have seen in sec.~\ref{constrsec} that in
case of a linearization around an $AdS$ geometry the composite
connections transform in the same way as the `fundamental'
connections -- and so imposing the constraints does not violate
the HS invariance --, it is not clear whether this is consistent
in general. In case it is not consistent, this would mean that
there are additional degrees of freedom associated to the torsion,
which necessarily need to be taken into account. Apart from that
we should point out that due to the way the torsion tensor enters
the Chern-Simons theory, there does not exist a 1.5--order
formalism.

One of the main difficulties in analyzing the non-linear dynamics
of the constructed HS theory in more detail is due to the fact
that the infinite-dimensional HS algebras are poorly understood.
Though these algebras are well-defined through the oscillator
realization described in sec.~\ref{sec:ahse}, the structure
constants, for instance, are not known in general. Further
research into this direction is required for a detailed analysis
of the interactions.

Once the dynamical content is known, it remains to be seen how the
different fields organize into HS multiplets. We first notice that
the basic field content of the Chern-Simons theory in $D=5$ does
not fit into multiplets of $\mhso(4,2)$, since the latter requires
in particular a massless scalar
\cite{Konshtein:1988yg,Engquist:2005yt}. This in turn is the
reason that the construction of HS actions \`a la
MacDowell-Mansouri entirely based on a HS gauge field are not
believed to be consistent to all orders \cite{Vasiliev:2001wa}.
However, our case is different, since there are no propagating HS
modes around $AdS_5$ and so there is no reason to expect that the
$D=5$ field content should organize in multiplets.\footnote{A
similar argument has been employed for supergravity in
\cite{Hohm:2005ui}.} Rather we found that the non-trivial HS
dynamics takes place on backgrounds which are not maximally
symmetric, as the $AdS_4$ solution. However, on this background
there will most likely be scalar and other excitations which are
the Kaluza-Klein modes originating from the off-diagonal
components of the various fields. Due to the HS invariance of the
full theory, these modes almost by construction will organize into
multiplets of $\mhso(3,2)$, and it would be very interesting to
see how this happens. In some sense the theory seems to prevent
itself from becoming inconsistent exactly by not having standard
dynamics around its most symmetric solution.

Let us finally note that the Chern-Simons theory in $D=11$ based
on (two copies of) the superalgebra $\mosp(1|32)$ has been
proposed as the non-perturbative definition of M-theory
\cite{Horava:1997dd}. As the latter should cover in particular the
infinite towers of massive HS states described by 10-dimensional
string theory, it is very tempting to conjecture that
$\mosp(1|32)$ has to be enhanced to a HS extension, thus giving
rise to Chern-Simons actions of the type considered here. In fact,
recently it has been argued that the three-dimensional
Chern-Simons theory based on a HS algebra is related to M-theory
for non-critical strings in $D=2$ via the background $AdS_2\times
S^1$ \cite{Horava:2007ds}. Similarly to the $AdS_4\times S^1$
solution discussed for the $D=5$ theory here, one might hope to
identify a non-topological 10-dimensional phase, which permits
flat Minkowski space and gives rise to massive HS states via
spontaneous symmetry breaking.

\subsection*{Acknowledgments}
For useful comments and discussions we would like to thank
D.~Francia, M.~Henneaux, C.~Iazeolla, M.A.~Vasiliev and especially
P.~Sundell.

This work has been supported by the European Union RTN network
MRTN-CT-2004-005104 {\it Constituents, Fundamental Forces and
Symmetries of the Universe} and the INTAS contract 03-51-6346 {\it
Strings, branes and higher-spin fields}. O.H. is supported by the
stichting FOM.

\begin{appendix}
\renewcommand{\theequation}{\Alph{section}.\arabic{equation}}

\section{Young Tableaux and
Projectors}\setcounter{equation}{0}\label{A} Here we give a brief
review of the technique of Young tableaux used in the main text.
As we are exclusively working with trace-full tensors, these
encode the irreducible representations of $GL(m)$, as opposed to
$SO(m)$ groups. For tensors with $AdS_D$ indices we have $m=D+1$,
while for the corresponding `Lorentz' tensors $m=D$.

A Young tableau consists of a certain number of rows of boxes,
where the number of boxes does not increase from top to bottom, as
for instance
 \bea\label{example}
  \yng(5,3,3,1)\;.
 \eea
It describes the symmetries of an irreducible $GL(m)$ tensor. For
the example (\ref{example}) it has the structure
$T_{a_1...a_5,b_1...b_3,c_1...c_3,d}$. As a matter of convention
we choose the symmetric basis, which means that the corresponding
tensors are completely symmetric in all row indices. Specifically,
the tensor $T$ above is completely symmetric in the sets of
indices $\{a_i\}$, $\{b_i\}$ and $\{c_i\}$, respectively. For
irreducibility the tensors have to satisfy the additional
condition that symmetrisation of all indices in a certain row with
any index corresponding to a box below that row gives zero. For
instance, in the example this implies
 \bea\label{irred}
  T_{a_1\ldots a_5,(b_1\ldots b_3,c_1)\ldots c_3,d} \ = \
  T_{a_1\ldots a_5,b_1\ldots b_3,(c_1\ldots c_3,d)}\ = \ 0\;, \quad {\rm
  etc.}\;,
 \eea
where ordinary brackets denote complete symmetrization of strength
one, as e.g.~$T_{(ab)}:=\frac{1}{2}(T_{ab}+T_{ba})$. Note that,
accordingly, a tensor $T_{a,b}$ in $\tiny{\yng(1,1)}$ is
antisymmetric, while in general no specific antisymmetrization
properties can be derived from the Young tableau.\footnote{It is,
however, possible to start with a different convention, in which
the \textit{antisymmetrization} properties, i.e.~the symmetries in
a column of boxes, are specified. In appendix~\ref{sec:appC} we
have to relate these two.} Moreover, one may check that for a
tensor in the `window' tableau $\tiny{\yng(2,2)}$\hspace{0.2em},
eq.~(\ref{irred}) implies the exchange property
$T_{ab,cd}=T_{cd,ab}$.

The language of Young tableaux is efficient in order to determine
the decomposition of tensor products into irreducible
representations. Specifically, in the tensor product of the
`vector' representation $\tiny{\yng(1)}$ with any Young tableau,
the irreducible parts are obtained by adding $\tiny{\yng(1)}$ to
the given tableau in all possible ways. For instance, the spin-3
frame field $\bein{e}{\mu}{ab}$ is a priori in the tensor product
 \bea\label{hooktens}
  {\small  \yng(1)} \hspace{0.5em} \otimes\hspace{0.5em}  {\small \yng(2) }
  \hspace{0.5em}=\hspace{0.5em}
  {\small \yng(3)}\hspace{0.5em}\oplus\hspace{0.5em} {\small
  \yng(2,1)}\;\;,
 \eea
i.e.~it contains the completely symmetric (physical) part and the
so-called `hook' diagram.

Finally we give the projectors onto the hook and window diagrams,
which we need in the main text, explicitly. The hook projector
reads on a general tensor with no a priori symmetries
 \bea
  (\mathbb{P}_{\tiny{\yng(2,1)}}X)_{abc}\ \equiv \
  (\mathbb{P}_{(2,1)}X)_{abc}\ \equiv\ X_{\langle abc\rangle}\ =\ \frac{1}{3}\left(2
  X_{(ab)c}-X_{(bc)a}-X_{(ca)b}\right)\;.
 \eea
Similarly,
 \begin{eqnarray}
  (\mathbb P_{\tiny{\yng(2,2)}} X)_{abcd} &\equiv& (\mathbb{P}_{(2,2)}X)_{abc}\ \equiv\ X_{\langle abcd\rangle} \\ \nonumber
  &=& \frac{1}{6}\left(2X_{(ab)(cd)}+2X_{(cd)(ab)}-X_{(cb)(ad)}-
  X_{(ad)(cb)}-X_{(ac)(bd)}-X_{(bd)(ac)}\right)\;.
 \eea
Analogous formulas hold in case of different index orderings, as
e.g.~hook projection according to indices $(ab,c)$ on a tensor
$X_{cab}$,
 \bea
  (\mathbb{P}_{\tiny{\yng(2,1)}}X)_{cab}\ = \
  \frac{1}{3}\left(2X_{c(ab)}-X_{a(bc)}-X_{b(ca)}\right)\;.
 \eea

\section{Proof of Cyclicity of the Trace} \la{sec:appB}

In this appendix we prove the assertion made in
sec.~\ref{sec:invtens} that the traces $\Tr_k$ in \eq{cyclicFG}
are cyclic in a general odd dimension $D=2n-1$.

For a generic element $F(M)$ in the enveloping algebra
$\cU[\mso(D-1,2)]$ the star product with $M_{AB}$ can be computed
by use of \eq{eopt}, \bea
\begin{split}
 & M_{AB}\star F\ =\ \Big(
M_{AB}+2M_{C[A}\del^C{}_{B]}+\frac23\big(M_{CD}\del_{[A}{}^C\del_{B]}{}^D-M_{C[A}\del_{B]}{}^D\del_{D}{}^C\Big)F\
, \\ & F \star M_{AB}\ =\  \Big(
M_{AB}-2M_{C[A}\del^C{}_{B]}+\frac23\big(M_{CD}\del_{[A}{}^C\del_{B]}{}^D-M_{C[A}\del_{B]}{}^D\del_{D}{}^C\big)\Big)F\
, \la{MstarF}
\end{split}
\eea which implies \bea \la{abcx}
[M_{AB},F(M)]_\star&=&4M_{C[A}\del^C{}_{B]}F(M)\ .
\eea
This
equation encodes the transformation of $F(M)$ under $M_{AB}$. In a
more mathematical language this states that the BCH star product
is strongly $\mso(D-1,2)$-invariant
\cite{gutt:1983,bieliavsky-2002}.

In order to prove the cyclicity of the trace we first show for a
{\it generic} monomial $F_\ell=F^{A_1B_1\ldots A_\ell
B_\ell}M_{A_1B_1}\cdots M_{A_\ell B_\ell}$ of degree $\ell$ \bea
\la{abcy}
\Tr_1([M_{AB},F_\ell]_\star)&=&\tr\big(\D([M_{AB},F_\ell]_\star)\big)\
= \ 0 \ . \eea To see this, we apply $\D$ to \eq{abcx}, whose
explicit evaluation gives
 \bea
 \begin{split}
  \tr\big(&\D([M_{AB},F_\ell]_\star)  \\
  &= 4\e^{A_1\ldots A_nB_1\ldots B_n}\sum_{r=0}^n
  \binom{n}{r}\tr\Big(\del_{A_1B_1}\cdots
  \del_{A_rB_r}M_{C[A|}\del_{A_{r+1}B_{r+1}}\cdots
  \del_{A_nB_n}\del^C{}_{|B]}F_\ell\Big)\\
  &=
  \tr\big([M_{AB},\D F_\ell]_\star\big)+4n\e^{A_1\ldots A_nB_1\ldots
  B_n}\tr\Big(\del_{A_1B_1}M_{C[A|}\del_{A_{2}B_{2}}\cdots
  \del_{A_nB_n}\del^C{}_{|B]}F_\ell\Big)\ .
 \end{split}
 \eea
The first term vanishes, which follows from \eq{abcx} and the fact
that `$\tr$' sets $M=0$. The second term can potentially be
nonzero when $\ell=n$. In this case it reduces to (after dropping
a constant multiplicative factor) $\e^{A_1\ldots A_nB_1\ldots
B_{n-1}}{}_{[A|} F_{A_1B_1\ldots A_{n-1}B_{n-1}A_n|B]}$, which
vanishes identically. To see this we use $F^{A_1B_1\ldots A_\ell
B_\ell}=-F^{A_1B_1\ldots B_mA_m\ldots A_\ell B_\ell}$, the
symmetry under exchange of any pair $(A_m,B_m)$ and
$(A_{m'},B_{m'})$ and finally the fact that antisymmetrization in
$2n+1$ indices vanishes identically for $\mso(2n)$,
 \bea \la{abcdd}
  \epsilon^{[A_1\ldots A_nB_1\ldots B_{n-1}A}F_{A_1B_1\ldots
  A_{n-1}B_{n-1}A_n}{}^{B]} \ =\ 0\; .
 \eea

Furthermore, one can show that for $k\ge2$ \bea \la{abcw}
\Tr_k([M_{AB},F_\ell]_\star)&=&\tr\big(\D^k([M_{AB},F_\ell]_\star)\big)\
= \ 0 \ , \eea by using identities similar to \eq{abcdd}, which
proves that $\Tr_k\big(F_\ell \star M_{AB}\big)$ is cyclic. By
using induction, the proof extends directly to
$\Tr_k\big(F_\ell\star G_{\ell'}\big)$ for an arbitrary monomial
$G_{\ell'}$. For this we expand $G_{\ell'}=\sum_mG_mM^m$, use that
$M^m=(M)^{\star m}+\sum_{m'<m}c_{m'}M^{\star m'}$ together with
associativity, and finally apply \eq{abcw} several times. For
instance, when $\ell'=2$ we find
$G_2=G^{A_1B_1A_2B_2}M_{A_1B_1}M_{A_2B_2}=G^{A_1B_1A_2B_2}(M_{A_1B_1}\star
M_{A_2B_2}-2\eta_{B_1A_2}M_{A_1B_2})$. The cyclicity of the last
term follows from the analysis above, and by repeatedly using
\eq{abcw} we have that
 \bea
  \begin{split}
  G^{A_1B_1A_2B_2}\Tr_k\Big(F_\ell\star M_{A_1B_1}\star
  M_{A_2B_2}\Big)\ &= \ G^{A_1B_1A_2B_2}\Tr_k\Big(M_{A_2B_2}\star
  F_\ell\star M_{A_1B_1}\Big)\\ \ &= \
  G^{A_1B_1A_2B_2}\Tr_k\Big(M_{A_1B_1}\star M_{A_2B_2}\star
  F_\ell\Big) \ .
  \end{split}
 \eea

\section{The Spin-3 Riemann Tensor} \la{sec:appC} Here we
summarize some relations for the spin-3 Riemann tensor, most
notably the Bianchi identities. (On flat space, a very clear
discussion of the spin-3 geometry in metric-like formulation can
be found in \cite{Damour:1987vm}, while aspects of a frame-like
formulation are given in
\cite{Vasiliev:1980as,Aragone:1981yn,Aragone:1988yx}.) For the
proof of the Bianchi identities it will be convenient to work in
form language, for which the tensors in (\ref{strength}) read
 \begin{eqnarray}
  T^{ab}&=&\bar{D}e^{ab}+\omega^{ab,c}\wedge \bar{e}_c\;, \qquad
  \nonumber
  T^{ab,c}\ =\ \bar{D}\omega^{ab,c}+3e^{\langle ab}\wedge
  \bar{e}^{c\rangle}+\omega^{ab,cd}\wedge\bar{e}_d\;, \\
  R^{ab,cd} &=& \bar{D}\omega^{ab,cd}+4\omega^{\langle
  ab,c}\wedge\bar{e}^{d\rangle}\;.
 \end{eqnarray}
After solving the torsion constraints the Bianchi identity follows
by application of $\bar{D}$ to the second torsion tensor,
 \bea\label{bianchi1}
  0\ =\ \bar{D}T^{ab,c}\ =\ (\bar{D}\omega^{ab,cd}+4\omega^{\langle
  ab,c}\wedge
  \bar{e}^{d\rangle})\wedge\bar{e}_d\ =\ R^{ab,cd}\wedge\bar{e}_d\;,
 \eea
where we used the first torsion constraint, $T^{ab}=0$, and the
relation
 \bea
  \bar{D}^2\omega^{ab,c}\ = \
  \bar{R}^{ad}\wedge\omega_d^{\hspace{0.5em}b,c}
  +\bar{R}^{bd}\wedge\omega_{\hspace{0.5em}d}^{a\hspace{0.5em},c}
  +\bar{R}^{cd}\wedge\omega_{\hspace{1em}d}^{ab,}\;,
 \eea
evaluated for the $AdS$ case (\ref{adscomm}). In components the
Bianchi identity (\ref{bianchi1}) reads
 \bea\label{bianchi2}
  R_{[\mu\nu\hspace{0.3em}\rho]\sigma,\lambda\delta}\ =\ 0\;,
 \eea
where we converted all indices into curved ones.

These identities can now be used to prove that all traces of the
Riemann tensor are algebraically related, or in other words, as in
the spin-2 case there is a unique Ricci tensor. First of all, the
symmetries of the fiber indices according to the window Young
tableau imply $R_{\mu\nu\hspace{0.3em}a(b,cd)}=0$, which in turn
shows that
 \bea
  R_{\mu\nu\hspace{0.3em}ab,\hspace{0.5em}c}^{\hspace{2.2em}c}\ =\
  -2R_{\mu\nu\hspace{0.3em}a\hspace{0.4em},bc}^{\hspace{1.7em}c}\
  =\ -2R_{\mu\nu\hspace{0.3em}b\hspace{0.4em},ac}^{\hspace{1.7em}c}\;,
 \eea
i.e.~there is a unique trace in the fiber indices. By virtue of
the Bianchi identity (\ref{bianchi2}) the trace in the fiber
indices can then be related to the trace between one space-time
and one fiber index:
 \bea\label{fibertr}
  R\undersym{_{\mu\lambda\hspace{0.2em}\nu\hspace{-0.2em}}{}^\lambda}{}_{,\rho\sigma}
  \ =\ \frac{1}{2}R_{\mu\nu
  \hspace{0.1em}\rho\sigma,}{}^\lambda{}_\lambda\;.
 \eea

We are now in a position to rigorously derive the field equations
used in the main text. First contracting (\ref{spin3eq}) with
$g^{\mu\rho}$ yields
 \bea
  (D-2)R_{\lambda\delta\hspace{0.9em}\nu,\hspace{0.5em}\sigma}
  ^{\hspace{1.2em}\lambda\hspace{0.7em}\delta}\ =\ 0\;,
 \eea
where we used (\ref{bianchi2}). This in turn implies that the
double traces of the Riemann tensor appearing in (\ref{spin3eq})
can be set to zero. The remaining terms can be simplified by
making repeated use of the Bianchi identity (\ref{bianchi2}) and
the symmetries of the window tableau:
 \begin{eqnarray} \begin{split}
  0 &\ =\  R_{\mu\lambda\hspace{0.8em}\nu,\rho\sigma}^{\hspace{1.2em}\lambda}-
  R_{\mu\lambda\hspace{0.8em}\sigma,\rho\nu}^{\hspace{1.2em}\lambda}
  +(\mu\leftrightarrow \sigma) \\
  &\ =\  R_{\rho\lambda\hspace{0.8em}\nu,\mu\sigma}^{\hspace{1.2em}\lambda}-
  R_{\rho\lambda\hspace{0.8em}\sigma,\mu\nu}^{\hspace{1.2em}\lambda}
  +(\mu\leftrightarrow \sigma) \\
  &\ =\ 4 R_{\rho\lambda\hspace{0.8em}\nu,\mu\sigma}^{\hspace{1.2em}\lambda}
  +R_{\rho\lambda\hspace{0.8em}\mu,\nu\sigma}^{\hspace{1.2em}\lambda}
  +R_{\rho\lambda\hspace{0.8em}\sigma,\nu\mu}^{\hspace{1.2em}\lambda}
  \\ &\ =\
  3R_{\rho\lambda\hspace{0.8em}\nu,\mu\sigma}^{\hspace{1.2em}\lambda}\;.
 \end{split} \end{eqnarray}
Here we used in the second line the Bianchi identity in
$\mu,\lambda,\rho$, in the third line the window symmetry of the
second term in $\mu,\nu,\sigma$ and finally the same symmetry in
the fourth line. These are the final equations of motion, which
basically state that the spin-3 Ricci tensor vanishes. In order to
clarify the information contained in this equation, we decompose
$R_{\lambda\mu\hspace{0.4em}\nu\rho,\sigma}^{\hspace{2.9em}\lambda}$
into its irreducible parts. A priori it can take values in the
Young tableaux
 \bea
  {\small  \yng(1)} \hspace{0.5em} \otimes\hspace{0.5em}  {\small \yng(2,1) }
  \hspace{0.5em}\ =\ \hspace{0.5em}
  {\small \yng(2,2)}\hspace{0.5em}\oplus\hspace{0.5em} {\small
  \yng(2,1,1)}\hspace{0.5em}\oplus\hspace{0.5em} {\small
  \yng(3,1)}\;\;.
 \eea
The origin of these different structures is that in the frame-like
formulation the Riemann tensor necessarily appears in a mixed
basis in the sense that the antisymmetric 2-form indices are on a
different footing as the frame indices. To compare with the
completely symmetric or completely antisymmetric basis used in the
metric-like formulation in \cite{deWit:1979pe} and
\cite{Damour:1987vm}, respectively, we have to impose these
symmetries, i.e.~we define
 \bea &&
  {\cal R}^{(a)}_{\mu\n;\r\s;\l\delta}\ =\
  R\undersym{{}_{\hspace{.02cm}\mu\r\hspace{.05cm}\n\phantom{\r}}}
  \hspace{-.2cm}\undersym{{}_{\hspace{.05cm}\l,\d\s}}\hspace{-1.0cm}\undersym{{}_{\phantom{TTTTT\hspace{.1cm}}}}\;, \qquad
  {\cal R}_{\mu\nu\rho,}^{(s)}{}^{\sigma\lambda\delta}\
  = \
  R_{(\mu}{}^{(\s}{}_{\nu\rho),}{}^{\lambda\delta)}\;.
 \eea

Since these are in definite Young tableaux (namely both in
$\tiny{\yng(3,3)}$, depending on the chosen conventions for
symmetrisation or antisymmetrisation properties), it is easily
seen that there is a unique trace. Explicitly one finds
 \bea
  {\cal R}^{(a)}_{\mu\n;\r;\s}\ =\ \frac{3}{8}R_{\mu\n
  \hspace{0.1em}\rho\s,}{}^\l{}_\l\;,
  \qquad
  {\cal R}^{(s)}_{\mu\nu\rho,\sigma}\ = \ \frac{1}{2}
  R_{\lambda(\mu\hspace{0.3em}\nu\rho),\sigma}^{\hspace{3.3em}\lambda}\;.
 \eea
With these relations it follows that this trace of ${\cal
R}^{(s)}$ is in $\tiny{\yng(3,1)}$, while its algebraically
related trace is in $\tiny{\yng(2,2)}$. Similarly, the trace of
${\cal R}^{(a)}$ takes values in ${\tiny \yng(3,1)}$, but
interpreted in the antisymmetric basis. To summarize, taking the
trace in the fiber indices of the Riemann tensor in the mixed
basis corresponds to the Ricci tensor in the completely
antisymmetric basis, while a trace between space-time and fiber
index corresponds to the Ricci tensor in the completely symmetric
basis.

\end{appendix}


\vskip 1cm

\begin{thebibliography}{99}

\bibitem{Gross:1988ue}
  D.~J.~Gross,
  ``High-energy symmetries of string theory,''
  Phys.\ Rev.\ Lett.\  {\bf 60} (1988) 1229.

\bibitem{Isberg:1993av}
  J.~Isberg, U.~Lindstrom, B.~Sundborg and G.~Theodoridis,
  ``Classical and quantized tensionless strings,''
  Nucl.\ Phys.\  B {\bf 411} (1994) 122
  [arXiv:hep-th/9307108].

\bibitem{Sundborg:2000wp}
  B.~Sundborg,
  ``Stringy gravity, interacting tensionless strings and massless higher
  spins,''
  Nucl.\ Phys.\ Proc.\ Suppl.\  {\bf 102} (2001) 113
  [arXiv:hep-th/0103247].

\bibitem{Bonelli:2003kh}
  G.~Bonelli,
  ``On the tensionless limit of bosonic strings, infinite symmetries and
  higher spins,''
  Nucl.\ Phys.\  B {\bf 669} (2003) 159
  [arXiv:hep-th/0305155].
  ``On the covariant quantization of tensionless bosonic strings in
  $AdS$ spacetime,''
  JHEP {\bf 0311} (2003) 028
  [arXiv:hep-th/0309222].

\bibitem{Sagnotti:2003qa}
  A.~Sagnotti and M.~Tsulaia,
  ``On higher spins and the tensionless limit of string theory,''
  Nucl.\ Phys.\  B {\bf 682} (2004) 83
  [arXiv:hep-th/0311257].

\bibitem{Engquist:2005yt}
  J.~Engquist and P.~Sundell,
  ``Brane partons and singleton strings,''
  Nucl.\ Phys.\  B {\bf 752} (2006) 206
  [arXiv:hep-th/0508124].

\bibitem{Sorokin:2004ie}
  D.~Sorokin,
  ``Introduction to the classical theory of higher spins,''
  AIP Conf.\ Proc.\  {\bf 767} (2005) 172
  [arXiv:hep-th/0405069].

\bibitem{Weinberg:1980kq}
  S.~Weinberg and E.~Witten,
  ``Limits on massless particles,''
  Phys.\ Lett.\  B {\bf 96} (1980) 59.

\bibitem{Aragone:1979bm}
  C.~Aragone and S.~Deser,
  ``Consistency problems of spin-2 gravity coupling,''
  Nuovo Cim.\  B {\bf 57} (1980) 33.

\bibitem{Coleman:1967ad}
  S.~R.~Coleman and J.~Mandula,
  ``All possible symmetries of the S matrix,''
  Phys.\ Rev.\  {\bf 159} (1967) 1251.

\bibitem{Bekaert:2005vh}
  X.~Bekaert, S.~Cnockaert, C.~Iazeolla and M.~A.~Vasiliev,
  ``Nonlinear higher spin theories in various dimensions,''
  Lectures given at Workshop on Higher Spin Gauge Theories, Brussels, Belgium, 12-14 May
  2004,
  arXiv:hep-th/0503128.

\bibitem{MacDowell:1977jt}
  S.~W.~MacDowell and F.~Mansouri,
  ``Unified geometric theory of gravity and supergravity,''
  Phys.\ Rev.\ Lett.\  {\bf 38} (1977) 739
  [Erratum-ibid.\  {\bf 38} (1977) 1376].

\bibitem{Fradkin:1986qy}
  E.~S.~Fradkin and M.~A.~Vasiliev,
  ``Cubic interaction in extended theories of massless higher spin fields,''
  Nucl.\ Phys.\ B {\bf 291} (1987) 141.

\bibitem{Vasiliev:2001wa}
  M.~A.~Vasiliev,
  ``Cubic interactions of bosonic higher spin gauge fields in $AdS_5$,''
  Nucl.\ Phys.\ B {\bf 616} (2001) 106
  [Erratum-ibid.\ B {\bf 652} (2003) 407]
  [arXiv:hep-th/0106200].

\bibitem{Vasiliev:2005zu}
  M.~A.~Vasiliev,
  ``Actions, charges and off-shell fields in the unfolded dynamics approach,''
  Int.\ J.\ Geom.\ Meth.\ Mod.\ Phys.\  {\bf 3} (2006) 37
  [arXiv:hep-th/0504090].

\bibitem{Vasiliev:1992gr}
  M.~A.~Vasiliev,
  ``Unfolded representation for relativistic equations in $(2+1)$ anti-De Sitter
  space,''
  Class.\ Quant.\ Grav.\  {\bf 11} (1994) 649.

\bibitem{Vasiliev:1989xz}
  M.~A.~Vasiliev,
  ``Triangle identity and free differential algebra of massless higher spins,''
  Nucl.\ Phys.\  B {\bf 324} (1989) 503.

\bibitem{Blencowe:1988gj}
  M.~P.~Blencowe,
  ``A consistent interacting massless higher spin field theory in $D = (2+1)$,''
  Class.\ Quant.\ Grav.\  {\bf 6} (1989) 443.

\bibitem{Fradkin:1989xt}
  E.~S.~Fradkin and V.~Y.~Linetsky,
  ``A superconformal theory of massless higher spin fields in $D = (2+1)$,''
  Mod.\ Phys.\ Lett.\  A {\bf 4} (1989) 731
  [Annals Phys.\  {\bf 198} (1990) 293].

\bibitem{Bergshoeff:1989ns}
  E.~Bergshoeff, M.~P.~Blencowe and K.~S.~Stelle,
  ``Area preserving diffeomorphisms and higher spin algebra,''
  Commun.\ Math.\ Phys.\  {\bf 128} (1990) 213.

\bibitem{Hohm:2005sc}
  O.~Hohm,
  ``On the infinite-dimensional spin-2 symmetries in Kaluza-Klein theories,''
  Phys.\ Rev.\ D {\bf 73} (2006) 044003
  [arXiv:hep-th/0511165].

\bibitem{Hohm:2006ud}
  O.~Hohm,
  ``Gauged diffeomorphisms and hidden symmetries in Kaluza-Klein theories,''
  Class.\ Quant.\ Grav.\ {\bf 24} (2007) 2825-2844.
  arXiv:hep-th/0611347.

\bibitem{Witten:1988hc}
  E.~Witten,
  ``(2+1)-dimensional gravity as an exactly soluble system,''
  Nucl.\ Phys.\ B {\bf 311} (1988) 46.

\bibitem{Chamseddine:1989nu}
  A.~H.~Chamseddine,
  ``Topological gauge theory of gravity in five-dimensions and all
  odd dimensions,''
  Phys.\ Lett.\ B {\bf 233} (1989) 291.

\bibitem{Chamseddine:1990gk}
  A.~H.~Chamseddine,
  ``Topological gravity and supergravity in various dimensions,''
  Nucl.\ Phys.\ B {\bf 346} (1990) 213.

\bibitem{Banados:1995mq}
  M.~Banados, L.~J.~Garay and M.~Henneaux,
  ``The local degrees of freedom of higher dimensional pure
  Chern-Simons theories,''
  Phys.\ Rev.\  D {\bf 53} (1996) 593
  [arXiv:hep-th/9506187].

\bibitem{Banados:1996yj}
  M.~Banados, L.~J.~Garay and M.~Henneaux,
  ``The dynamical structure of higher dimensional Chern-Simons theory,''
  Nucl.\ Phys.\ B {\bf 476} (1996) 611
  [arXiv:hep-th/9605159].

\bibitem{Fronsdal:1978rb}
  C.~Fr\o nsdal,
  ``Massless fields with integer spin,''
  Phys.\ Rev.\  D {\bf 18} (1978) 3624.

\bibitem{Fronsdal:1978vb}
  C.~Fr\o nsdal,
  ``Singletons and massless, integral spin fields on de Sitter space,''
  Phys.\ Rev.\  D {\bf 20} (1979) 848.

\bibitem{deWit:1979pe}
  B.~de Wit and D.~Z.~Freedman,
  ``Systematics of higher spin gauge fields,''
  Phys.\ Rev.\  D {\bf 21} (1980) 358.

\bibitem{Fradkin:1986ka}
  E.~S.~Fradkin and M.~A.~Vasiliev,
  ``Candidate to the role of higher spin symmetry,''
  Annals Phys.\  {\bf 177} (1987) 63.

\bibitem{Vasiliev:1980as}
  M.~A.~Vasiliev,
  `` `Gauge' form of description of massless fields with arbitrary spin. (In
  Russian),''
  Yad.\ Fiz.\  {\bf 32} (1980) 855.

\bibitem{Vasiliev:2003ev}
  M.~A.~Vasiliev,
  ``Nonlinear equations for symmetric massless higher spin fields in
  $(A)dS(d)$,''
  Phys.\ Lett.\  B {\bf 567} (2003) 139
  [arXiv:hep-th/0304049].

\bibitem{Sagnotti:2005ns}
  A.~Sagnotti, E.~Sezgin and P.~Sundell,
  ``On higher spins with a strong $Sp(2,R)$ condition,''
  arXiv:hep-th/0501156.

\bibitem{Sezgin:2001zs}
  E.~Sezgin and P.~Sundell,
  ``Doubletons and $5D$ higher spin gauge theory,''
  JHEP {\bf 0109} (2001) 036
  [arXiv:hep-th/0105001].

\bibitem{bieliavsky-2002}
  P.~Bieliavsky, M.~Bordemann, S.~Gutt and S.~Waldmann,
  ``Traces for star products on the dual of a Lie algebra,''
  arXiv.org:math/0202126.

\bibitem{gutt:1983}
 S.~Gutt, ``An explicit-product on the cotangent bundle of a Lie
group,'' Lett. Math. Phys. {\bf 7} (1983) 249.

\bibitem{Madore:2000en}
  J.~Madore, S.~Schraml, P.~Schupp and J.~Wess,
  ``Gauge theory on noncommutative spaces,''
  Eur.\ Phys.\ J.\  C {\bf 16} (2000) 161
  [arXiv:hep-th/0001203].

\bibitem{Jurco:2000ja}
  B.~Jurco, S.~Schraml, P.~Schupp and J.~Wess,
  ``Enveloping algebra valued gauge transformations for non-Abelian gauge
  groups on non-commutative spaces,''
  Eur.\ Phys.\ J.\  C {\bf 17} (2000) 521
  [arXiv:hep-th/0006246].

\bibitem{Vasiliev:1986qx}
  M.~A.~Vasiliev,
  ``Extended higher spin superalgbras and their realizations in terms
  of quantum operators,''
  Fortsch.\ Phys.\  {\bf 36} (1988) 33.

\bibitem{Pinczon}
  G.~Pinczon and R.~Ushirobira,
  ``Supertrace and superquadratic Lie structure on the Weyl algebra, and
  applications to formal inverse Weyl transform ,''
  Lett.\ Math.\ Phys.\ {\bf 74} (2005) 263.

\bibitem{Zanelli:2002qm}
  J.~Zanelli,
  ``(Super)-gravities beyond $4$ dimensions,''
  arXiv:hep-th/0206169.

\bibitem{Fuchs:1997jv}
  J.~Fuchs and C.~Schweigert,
  ``Symmetries, Lie algebras and representations: A graduate course for
  physicists,'' Cambridge University Press, 1997.


\bibitem{Vasiliev:1986td}
  M.~A.~Vasiliev,
  ``Free massless fields of arbitrary spin in the de Sitter space and initial
  data for a higher spin superalgebra,''
  Fortsch.\ Phys.\  {\bf 35} (1987) 741
  [Yad.\ Fiz.\  {\bf 45} (1987) 1784].

\bibitem{Lopatin:1987hz}
  V.~E.~Lopatin and M.~A.~Vasiliev,
  ``Free massless bosonic fields of arbitrary spin in $d$-dimensional
  de Sitter space,''
  Mod.\ Phys.\ Lett.\  A {\bf 3} (1988) 257.

\bibitem{Francia:2002aa}
  D.~Francia and A.~Sagnotti,
  ``Free geometric equations for higher spins,''
  Phys.\ Lett.\  B {\bf 543} (2002) 303
  [arXiv:hep-th/0207002].

\bibitem{Francia:2002pt}
  D.~Francia and A.~Sagnotti,
  ``On the geometry of higher-spin gauge fields,''
  Class.\ Quant.\ Grav.\  {\bf 20} (2003) S473
  [arXiv:hep-th/0212185].

\bibitem{Francia:2006hp}
  D.~Francia and A.~Sagnotti,
  ``Higher-spin geometry and string theory,''
  J.\ Phys.\ Conf.\ Ser.\  {\bf 33} (2006) 57
  [arXiv:hep-th/0601199].

\bibitem{Damour:1987vm}
  T.~Damour and S.~Deser,
  ``'Geometry' Of Spin $3$ Gauge Theories,''
  Annales Poincare Phys.\ Theor.\  {\bf 47} (1987) 277.

\bibitem{Bekaert:2003az}
  X.~Bekaert and N.~Boulanger,
  ``On geometric equations and duality for free higher spins,''
  Phys.\ Lett.\  B {\bf 561} (2003) 183
  [arXiv:hep-th/0301243].

\bibitem{Bekaert:2006ix}
  X.~Bekaert and N.~Boulanger,
  ``Tensor gauge fields in arbitrary representations of $GL(D,R)$. II: Quadratic
  actions,''
  Commun.\ Math.\ Phys.\  {\bf 271} (2007) 723
  [arXiv:hep-th/0606198].

\bibitem{Bandos:2005mb}
  I.~Bandos, X.~Bekaert, J.~A.~de Azcarraga, D.~Sorokin and M.~Tsulaia,
  ``Dynamics of higher spin fields and tensorial space,''
  JHEP {\bf 0505} (2005) 031
  [arXiv:hep-th/0501113].


\bibitem{Banados:1997df}
  M.~Banados,
  ``Constant curvature black holes,''
  Phys.\ Rev.\  D {\bf 57} (1998) 1068
  [arXiv:gr-qc/9703040].

\bibitem{Edelstein:2006se}
  J.~D.~Edelstein, M.~Hassaine, R.~Troncoso and J.~Zanelli,
  ``Lie-algebra expansions, Chern-Simons theories and the Einstein-Hilbert
  lagrangian,''
  Phys.\ Lett.\  B {\bf 640} (2006) 278
  [arXiv:hep-th/0605174].

\bibitem{Sezgin:2005pv}
  E.~Sezgin and P.~Sundell,
  ``An exact solution of $4D$ higher-spin gauge theory,''
  Nucl.\ Phys.\  B {\bf 762} (2007) 1
  [arXiv:hep-th/0508158].

\bibitem{Didenko:2006zd}
  V.~E.~Didenko, A.~S.~Matveev and M.~A.~Vasiliev,
  ``BTZ black hole as solution of $3d$ higher spin gauge theory,''
  arXiv:hep-th/0612161.

\bibitem{Konshtein:1988yg}
  S.~E.~Konstein and M.~A.~Vasiliev,
  ``Massless representations and admissability condition for higher
  spin superalgebras,''
  Nucl.\ Phys.\  B {\bf 312} (1989) 402.


\bibitem{Hohm:2005ui}
  O.~Hohm and H.~Samtleben,
  ``Effective actions for massive Kaluza-Klein states on $AdS_3\times S^3
  \times S^3$,'' JHEP {\bf 0505} (2005) 027
  [arXiv:hep-th/0503088].

\bibitem{Aragone:1981yn}
  C.~Aragone and H.~La Roche,
  ``Massless second order tetradic spin 3 fields and higher helicity bosons,''
  Nuovo Cim.\  A {\bf 72} (1982) 149.

\bibitem{Aragone:1988yx}
  C.~Aragone, S.~Deser and Z.~Yang,
  ``Massive higher spin from dimensional reduction of gauge fields,''
  Annals Phys.\  {\bf 179} (1987) 76.

\bibitem{Horava:1997dd}
  P.~Horava,
  ``M-theory as a holographic field theory,''
  Phys.\ Rev.\  D {\bf 59} (1999) 046004
  [arXiv:hep-th/9712130].

\bibitem{Horava:2007ds}
  P.~Horava and C.~A.~Keeler,
  ``Strings on $AdS_2$ and the high-energy limit of noncritical M-Theory,''
  arXiv:0704.2230 [hep-th].

\end{thebibliography}
\end{document}